\documentclass[12pt]{article}
\usepackage{amssymb,amsmath,bm}
\usepackage{epsf}
\usepackage{epsfig}
\usepackage{cite}
\usepackage{bbm}
\usepackage{color}

\def\dfrac#1#2{{\displaystyle {#1 \over #2}}}

\def\simge{\mathrel{\rlap{\raise 0.511ex \hbox{$>$}}{\lower 0.511ex \hbox{$\sim$}}}}
\def\simle{\mathrel{\rlap{\raise 0.511ex \hbox{$<$}}{\lower 0.511ex \hbox{$\sim$}}}}
\def\slash#1{\setbox0=\hbox{$#1$}\dimen0=\wd0
      \setbox1=\hbox{/} \dimen1=\wd1 \ifdim\dimen0>\dimen1
      \rlap{\hbox to \dimen0{\hfil/\hfil}} #1                        \else
      \rlap{\hbox to \dimen1{\hfil$#1$\hfil}}
      /   \fi}

\newcommand{\lsim}{
\mathrel{\hbox{\rlap{\hbox{\lower4pt\hbox{$\sim$}}}\hbox{$<$}}}}

\newcommand{\gsim}{
\mathrel{\hbox{\rlap{\hbox{\lower4pt\hbox{$\sim$}}}\hbox{$>$}}}}

\def\eps{\varepsilon}

\newcommand{\tev}{\, {\rm TeV}}
\newcommand{\gev}{\, {\rm GeV}}
\newcommand{\mev}{\, {\rm MeV}}
\newcommand{\kev}{\, {\rm keV}}

\newcommand{\Heff}{{\cal H}_{\rm eff}}

\allowdisplaybreaks[1]

\newcommand{\be}{\begin{equation}}
\newcommand{\ee}{\end{equation}}
\newcommand{\bea}{\begin{eqnarray}}
\newcommand{\eea}{\end{eqnarray}}
\newcommand{\nn}{\nonumber}
\newcommand{\bi}{\begin{itemize}}
\newcommand{\ei}{\end{itemize}}
\newcommand{\ord}{{\cal O}}

\def\klpn{K_{L}\rightarrow\pi^0\nu\bar\nu}

\newcommand{\newsection}[1]{\section{#1}\setcounter{equation}{0}}

\begin{document}
\begin{titlepage}
\vspace*{-0.5truecm}

\begin{flushright}
TUM-HEP-657/07\\
MPP-2007-17
\end{flushright}

\vfill
\begin{center}
\boldmath

{\Large\textbf{Charged Lepton Flavour Violation and $\bm{(g-2)_\mu}$\vspace{-0.15cm}\\
 in the Littlest Higgs Model with T-Parity: \vspace{0.15cm}\\
 a clear Distinction from Supersymmetry}}

\unboldmath
\end{center}

\vspace{0.4truecm}

\begin{center}
{\large\bf Monika Blanke$^{a,b}$, Andrzej J.~Buras$^a$, Bj{\"o}rn Duling$^a$,\\
Anton Poschenrieder$^a$ and Cecilia Tarantino$^a$}
\vspace{0.4truecm}

 {\sl $^a$Physik Department, Technische Universit\"at M\"unchen,
D-85748 Garching, Germany}
\vspace{0.2cm}

 {\sl $^b$Max-Planck-Institut f{\"u}r Physik (Werner-Heisenberg-Institut), \\
D-80805 M{\"u}nchen, Germany}

\end{center}

\vspace{1cm}
\begin{abstract}
\vspace{0.2cm}
\noindent 
We calculate the rates for the charged lepton flavour violating decays $\ell_i\to\ell_j\gamma$, $\tau\to\ell\pi$, $\tau\to\ell\eta$, $\tau\to\ell\eta'$, $\mu^-\to e^- e^+ e^-$, the six three body leptonic decays $\tau^-\to \ell_i^-\ell_j^+\ell_k^-$ and the rate for $\mu-e$ conversion in nuclei in the Littlest Higgs 
model 
with T-Parity (LHT). We also calculate the rates for $K_{L,S}\to\mu e$, $K_{L,S}\to\pi^0\mu e$ and $B_{d,s}\to\ell_i\ell_j$. We find that the relative effects of mirror leptons in these transitions are by many orders of magnitude larger than analogous mirror quark effects  in rare $K$ and $B$ decays analyzed recently. In particular, in order to suppress the $\mu\to e \gamma$ and $\mu^-\to e^-e^+e^-$ decay rates and the $\mu-e$ conversion rate below the experimental upper bounds, the relevant mixing matrix in the mirror lepton sector $V_{H\ell}$ must be rather hierarchical, unless the spectrum of mirror leptons is quasi-degenerate. 
We find that the pattern of the LFV branching ratios in the LHT model differs significantly from the one encountered in the MSSM, allowing in a transparent manner to distinguish these two models with the help of LFV processes.
 We also calculate $(g-2)_\mu$ and find the new contributions to $a_\mu$ below $1\cdot 10^{-10}$ and consequently negligible. We compare our results with those present in the literature.
\end{abstract}
\vfill\vfill
\end{titlepage}

\thispagestyle{empty}

\begin{center}
{\Large\bf Note added}
\end{center}

\noindent
An additional contribution to the $Z$ penguin in the Littlest Higgs model with T-parity has been pointed out in \cite{Goto:2008fj,delAguila:2008zu}, which has been overlooked in the present analysis. This contribution leads to the cancellation of the left-over quadratic divergence in the calculation of some rare decay amplitudes. Instead of presenting separate errata to the present work and our papers \cite{Blanke:2006eb,Blanke:2007ee,Blanke:2007wr,Blanke:2008ac} partially affected by this omission, we have presented a corrected and updated analysis of flavour changing neutral current processes in the Littlest Higgs model with T-parity in \cite{Blanke:2009am}.

\newpage

\setcounter{page}{1}
\pagenumbering{arabic}

\newsection{Introduction}
\label{sec:intro}

Little Higgs models \cite{oldLH,LHreview,l2h} offer an alternative route
to the solution of the little hierarchy problem. One of the most attractive models of
this class is the Littlest Higgs model with T-parity (LHT) \cite{tparity},
where the discrete symmetry forbids tree-level corrections to electroweak
  observables, thus weakening the electroweak precision constraints \cite{mH}.
In this model, the new gauge bosons, 
fermions and scalars are sufficiently light to be 
discovered at LHC and there is a dark matter candidate \cite{LHTphen}. Moreover, 
the flavour structure of the LHT model is richer than the one of the 
Standard Model (SM), mainly due to the presence of three doublets 
of mirror quarks and three doublets of mirror leptons and their 
weak interactions with the ordinary quarks and leptons.

Recently, in two detailed analyses, we have investigated  in the LHT model $\Delta F=2$ \cite{BBPTUW} and $\Delta F=1$ \cite{Blanke:2006eb} flavour changing neutral current (FCNC) processes, like particle-antiparticle mixings, $B\to X_s\gamma$, $B\to X_s\ell^+\ell^-$ and rare $K$ and $B$ decays. The first analysis of particle-antiparticle mixing in this model was presented in \cite{Hubisz} and the FCNC processes in the LH model without T-parity have been presented in \cite{BPU,Indian,BPU2,BsgLH,BPUB}.

The most relevant messages of the phenomenological analyses in \cite{BBPTUW,Blanke:2006eb,BPU,BPUB} are:
\bi
\item
In the LH model without T-parity, which belongs to the class of models with
constrained minimal flavour violation (CMFV) \cite{CMFV,mfvlectures}, the new
physics {(NP)} effects are small as the {NP} scale $f$ is required to be above $2-3\tev$ in order to satisfy the electroweak precision constraints.
\item
In the LHT model, which is not stringently constrained by the latter precision tests and contains new flavour and CP-violating interactions, large departures from the SM predictions are found, in particular for CP-violating observables that are strongly suppressed in the SM. These are first of all the branching ratio for $\klpn$ and the CP asymmetry $S_{\psi\phi}$ in the $B_s\to \psi\phi$ decay, but also $Br(K_L\to\pi^0\ell^+\ell^-)$ and $Br(K^+\to\pi^+\nu\bar\nu)$. Smaller, but still significant, effects have been found in rare $B_{s,d}$ decays and $\Delta M_{s,d}$.
\item
The presence of left-over divergences in $\Delta F=1$ processes, that signals
some sensitivity to the ultraviolet {(UV)} completion of the theory,
introduces some theoretical uncertainty in the evaluation of the relevant
branching ratios both in the LH model \cite{BPUB} and the LHT model
\cite{Blanke:2006eb}. {On the other hand}, $\Delta F=2$ processes and the $B\to X_s\gamma$ decay are free from these divergences.
\ei

Now, it is well known that in the SM the FCNC processes in the lepton sector, like $\ell_i\to\ell_j\gamma$ and $\mu^-\to e^-e^+e^-$,
are very strongly suppressed due to tiny neutrino masses. In particular, the
branching ratio for $\mu\to e \gamma$ in the SM amounts to at most $10^{-54}$,
to be compared with the present experimental upper bound, $1.2\cdot 10^{-11}$ \cite{muegamma},
and with the one that will be available within the next two years,
$\sim 10^{-13}$ \cite{megexp}. 
Results close to the SM predictions are expected within the LH model without T-parity, where the lepton sector is identical to the one of the SM and the additional $\ord(v^2/f^2)$ corrections have only minor impact on this result. {Similarly the new effects on $(g-2)_\mu$ turn out to be small \cite{LHg-2}.}

A very different situation is to be expected in the LHT model, where the
presence of new flavour violating interactions and of mirror leptons with
masses of order $1\tev$ can  change the SM expectations up to 45
orders of magnitude, bringing the relevant branching ratios for lepton flavour
violating (LFV) processes close to the bounds available presently or in the near future. Indeed in two recent interesting {papers \cite{Goyal,IndianLFV}, it has been} pointed out that very large enhancements of the branching ratios for $\ell_i\to\ell_j\gamma$ and $\tau\to\mu\pi$ are possible within the LHT model.

The main goal of our paper is a new analysis of $\ell_i\to\ell_j\gamma$,
$\tau\to\mu\pi$, and of other LFV processes not considered in
\cite{Goyal,IndianLFV}, with the aim to {find the pattern of LFV in this model and to} constrain the mass spectrum of mirror
leptons and the new weak mixing matrix in the lepton sector $V_{H\ell}$, that
in addition to three mixing angles contains three CP-violating
phases\footnote{A detailed analysis of the number of phases in the mixing
  matrices in the LHT model has recently been presented {in \cite{SHORT}}.}. In particular we have calculated the rates for $\mu^-\to e^-e^+e^-$ and the six three body leptonic $\tau$ decays $\tau^-\to\ell_i^-\ell_j^+\ell_k^-$, as well as the $\mu-e$ conversion rate in nuclei. We have also calculated the rates for $K_{L,S}\to\mu e$, $K_{L,S}\to\pi^0\mu e$ and $B_{d,s}\to\ell_i\ell_j$ that are sensitive to flavour violation both in the mirror quark and mirror lepton sectors. Finally we calculated $(g-2)_\mu$ that has also been considered in \cite{Goyal,IndianLFV}.

Our analysis confirms the findings of \cite{Goyal,IndianLFV} at the
qualitative level: the impact of mirror leptons on the charged LFV processes
$\ell_i\to\ell_j\gamma$ and $\tau\to\mu\pi$ can be spectacular while the
impact on $(g-2)_\mu$ is small, {although our analytical expressions
differ from the ones presented in \cite{Goyal,IndianLFV}. Moreover, our
  numerical analysis includes also other LFV processes, not considered in
  \cite{Goyal,IndianLFV}, where very large effects turn out to be possible.}

While the fact that in the LHT model several LFV branching ratios can reach
their present experimental upper bounds is certainly interesting, it depends
sensitively on the parameters of the model. {One of the most important
  results of the present paper is the identification of
  correlations between various branching ratios that on the one hand are less
  parameter dependent and on the other hand, and more importantly, differ
  significantly from corresponding correlations in the Minimal Supersymmetric Standard Model (MSSM) discussed in \cite{Ellis,Brignole,Herrero,Paradisi}. The origin of this
  difference is that} the dominance of the dipole operators in the decays in
question present in the MSSM is replaced in the LHT model by the dominance of
$Z^0$-penguin and box diagram contributions with the dipole operators playing
now a negligible role. {As a consequence}, LFV processes can help to distinguish these two models.

A detailed analysis of LFV in the LHT model is also motivated by the prospects
in the measurements of LFV processes with much higher sensitivity than
presently available. In particular the MEG experiment at PSI \cite{megexp}
should be able to test $Br(\mu\to e\gamma)$ at the level of
$\ord(10^{-13}-10^{-14})$, and the Super Flavour Factory \cite{SuperB} is
planned to reach a sensitivity for $Br(\tau\to\mu\gamma)$ of at least
$\ord(10^{-9})$. {The planned accuracy of SuperKEKB of $\ord(10^{-8})$ for
  $\tau\to\mu\gamma$ is also of great interest.} Very important will also be an improved upper bound on $\mu-e$ conversion in Ti. In this context the dedicated J-PARC experiment PRISM/PRIME \cite{J-PARK} should reach the sensitivity of $\ord(10^{-18})$, i.\,e. an improvement by six orders of magnitude relative to the present upper bound from SINDRUM II at PSI \cite{mue-conv_bound}.

Our paper is organized as follows. In Section \ref{sec:model} we briefly summarize those ingredients of the LHT model that are of relevance for our analysis. Section \ref{sec:liljgamma} is devoted to the decays $\ell_i\to\ell_j\gamma$ with particular attention to $\mu\to e\gamma$, for which a new stringent experimental upper bound should be available in the coming years. In Section \ref{sec:semilep} we calculate the branching ratio for $\tau\to\mu\pi$ and other semi-leptonic $\tau$ decays for which improved upper bounds are available from Belle. In Section \ref{sec:mueee} we analyze the decays $\mu^-\to e^- e^+ e^-$, $\tau^-\to\mu^-\mu^+\mu^-$ and $\tau^-\to e^- e^+ e^-$. In Section \ref{sec:mueconv} we calculate the $\mu-e$ conversion rate in nuclei, and in Section \ref{sec:KLmue} the decays $K_{L,S}\to\mu e$ and $K_{L,S}\to\pi^0\mu e$. In Section \ref{sec:Bsd} we give the results for $B_{d,s}\to \mu e,\,\tau e,\, \tau\mu$ and in Sections \ref{sec:teme} and \ref{sec:tmee} for $\tau^-\to e^-\mu^+e^-,\,\mu^-e^+\mu^-,\,\mu^-e^+e^-,\,e^-\mu^+\mu^-$.
In Section \ref{sec:g-2} we calculate $(g-2)_\mu$. A detailed numerical analysis of all these processes is presented in Section \ref{sec:num}. In Section \ref{sec:corr} we analyze various correlations between LFV branching ratios and compare them with the MSSM results in \cite{Ellis,Brignole,Herrero,Paradisi}.
Finally, in Section \ref{sec:concl} we conclude our paper with a list of messages from our analysis and with a brief outlook. Few technical details are relegated to the appendices.

\newsection{The LHT Model and its Lepton Sector}\label{sec:model}

A detailed description of the LHT model and the relevant Feynman rules can be found  in
\cite{Blanke:2006eb}. Here we just want to state briefly the
ingredients needed for the present analysis.

\subsection{Gauge Boson Sector}
\label{subsec:2.1}

The T-even electroweak gauge boson sector  consists only of the SM
electroweak gauge bosons $W^\pm_L$, $Z_L$ and $A_L$.

The T-odd gauge boson sector consists of three heavy
``partners'' of the SM gauge bosons
\begin{equation}\label{2.3}
W_H^\pm\,,\qquad Z_H\,,\qquad A_H\,,
\end{equation}
with masses given to lowest order in $v/f$ by
\begin{equation}\label{2.4}
M_{W_H}=gf\,,\qquad M_{Z_H}=gf\,,\qquad
M_{A_H}=\frac{g'f}{\sqrt{5}}\,.
\end{equation}
All three gauge bosons will be present in our analysis. Note
that
\begin{equation}\label{2.4a}
M_{A_H}=\frac{\tan{\theta_W}}{\sqrt{5}}M_{W_H}\simeq\frac{M_{W_H}}{4.1}\,,
\end{equation}
where $\theta_W$ is the weak mixing angle.

\subsection{Fermion Sector}
\label{subsec:2.2}

The T-even sector of the LHT {model} contains just the SM fermions and the heavy top partner $T_+$. Due to the smallness of neutrino masses, the T-even contributions to LFV processes can be neglected with respect to the T-odd sector. We comment on the issue of neutrino masses in the LHT model in Appendix \ref{sec:app1}.

The T-odd fermion sector~\cite{mirror} consists of three
generations of mirror quarks and leptons with vectorial
couplings under $SU(2)_L\times U(1)_Y$. In this paper, except for $K_{L,S}\to\mu e$, $K_{L,S}\to\pi^0\mu e$, $B_{d,s}\to\ell_i\ell_j$ and $\tau\to\ell\pi,\ell\eta,\ell\eta'$, only  mirror leptons are
relevant. We will denote them by 
  \begin{equation}\label{2.6}
\begin{pmatrix} \nu^1_{H}\\\ell^1_{H} \end{pmatrix}\,,\qquad
\begin{pmatrix} \nu^2_{H}\\\ell^2_{H} \end{pmatrix}\,,\qquad
\begin{pmatrix} \nu^3_{H}\\\ell^3_{H} \end{pmatrix}\,,
\end{equation}
with their masses satisfying to first order in $v/f$
\begin{equation}\label{2.7}
m^\nu_{H1}=m^\ell_{H1}\,,\qquad m^\nu_{H2}=m^\ell_{H2}\,,\qquad
m^\nu_{H3}=m^\ell_{H3}\,.
\end{equation}

\subsection{Weak Mixing in the Mirror Lepton Sector}
\label{subsec:2.4}

As discussed in detail in~\cite{Hubisz}, one of the important
ingredients of the mirror sector is the existence of four CKM-like \cite{ckm} unitary
mixing matrices, two for mirror quarks and two
for mirror leptons:
\begin{equation}\label{2.10}
V_{Hu}\,,\quad V_{Hd}\,,\qquad V_{H\ell}\,,\quad V_{H\nu}\,.
\end{equation}
They satisfy\footnote{Note that it is $V_\text{CKM}$ but $V_\text{PMNS}^\dagger$ appearing on the r.\,h.\,s., as the PMNS matrix is defined through neutrino mixing, while the CKM matrix is defined through mixing in the down-type quark sector.}
\begin{equation}\label{2.11}
V_{Hu}^\dagger V_{Hd}=V_\text{CKM}\,,\qquad V_{H\nu}^\dagger
V_{H\ell}=V_\text{PMNS}^\dagger\,,
\end{equation}
where in $V_\text{PMNS}$~\cite{pmns} the Majorana phases are set
to zero as no Majorana mass term has
been introduced {for right-handed neutrinos}. The mirror mixing
matrices in \eqref{2.10} parameterize flavour violating
interactions between SM fermions and mirror fermions
that are mediated by the heavy gauge bosons $W_H^\pm$, $Z_H$ and $A_H$.
The notation in \eqref{2.10} indicates which of the light fermions
of a given electric charge participates in the interaction.

Thus $V_{H\ell}$, the most important mixing matrix in the present
paper, parameterizes the interactions of light charged leptons with
mirror neutrinos, mediated by $W_H^\pm$, and with mirror charged leptons,
mediated by $Z_H$ and $A_H$. Feynman rules for these interactions
can be found in \cite{Blanke:2006eb}.  $V_{H\nu}$ parameterizes, on the other hand, the
interactions of light neutrinos with mirror leptons. 

In the course of our analysis of charged LFV
decays it will be useful to introduce the following
quantities ($i=1,2,3$):
\begin{equation}\label{eq:chi}
\chi_i^{(\mu e)}=V^{*ie}_{H\ell}V^{i\mu}_{H\ell}\,,\qquad
\chi_i^{(\tau e)}=V^{*ie}_{H\ell}V^{i\tau}_{H\ell}\,,\qquad
\chi_i^{(\tau\mu)}=V^{*i\mu}_{H\ell}V^{i\tau}_{H\ell}\,,
\end{equation}
that govern $\mu\to e$, $\tau\to e$ and $\tau\to\mu$ transitions, respectively.

We also recall the analogous quantities in the mirror quark sector $(i=1,2,3)$
\be
\xi_i^{(K)}=V^{*is}_{Hd}V^{id}_{Hd}\,,\qquad
\xi_i^{(d)}=V^{*ib}_{Hd}V^{id}_{Hd}\,,\qquad
\xi_i^{(s)}=V^{*ib}_{Hd}V^{is}_{Hd}\,,
\ee
that we will  need for the analysis of the decays $K_{L,S}\to\mu e$, $K_{L,S}\to\pi^0\mu e$  and $B_{d,s}\to\ell_i\ell_j$.

Following  \cite{SHORT}, we parameterize $V_{H\ell}$ in terms of three mixing angles $\theta_{ij}^\ell$ and three complex phases $\delta_{ij}^\ell$
 as a product of three rotations, and
introducing a complex phase in each of them\footnote{Note that the two additional phases in $V_{H\ell}$ have nothing to do with the possible Majorana nature of neutrinos.}, thus obtaining
\be
\addtolength{\arraycolsep}{3pt}
V_{H\ell}= \begin{pmatrix}
1 & 0 & 0\\
0 & c_{23}^\ell & s_{23}^\ell e^{- i\delta^\ell_{23}}\\
0 & -s_{23}^\ell e^{i\delta^\ell_{23}} & c_{23}^\ell\\
\end{pmatrix}\,\cdot
 \begin{pmatrix}
c_{13}^\ell & 0 & s_{13}^\ell e^{- i\delta^\ell_{13}}\\
0 & 1 & 0\\
-s_{13}^\ell e^{ i\delta^\ell_{13}} & 0 & c_{13}^\ell\\
\end{pmatrix}\,\cdot
 \begin{pmatrix}
c_{12}^\ell & s_{12}^\ell e^{- i\delta^\ell_{12}} & 0\\
-s_{12}^\ell e^{i\delta^\ell_{12}} & c_{12}^\ell & 0\\
0 & 0 & 1\\
\end{pmatrix}\ee
Performing the product one obtains the expression
\be\label{2.12a}
\addtolength{\arraycolsep}{3pt}
V_{H\ell}= \begin{pmatrix}
c_{12}^\ell c_{13}^\ell & s_{12}^\ell c_{13}^\ell e^{-i\delta^\ell_{12}}& s_{13}^\ell e^{-i\delta^\ell_{13}}\\
-s_{12}^\ell c_{23}^\ell e^{i\delta^\ell_{12}}-c_{12}^\ell s_{23}^\ell s_{13}^\ell e^{i(\delta^\ell_{13}-\delta^\ell_{23})} &
c_{12}^\ell c_{23}^\ell-s_{12}^\ell s_{23}^\ell s_{13}^\ell e^{i(\delta^\ell_{13}-\delta^\ell_{12}-\delta^\ell_{23})} &
s_{23}^\ell c_{13}^\ell e^{-i\delta^\ell_{23}}\\
s_{12}^\ell s_{23}^\ell e^{i(\delta^\ell_{12}+\delta^\ell_{23})}-c_{12}^\ell c_{23}^\ell s_{13}^\ell e^{i\delta^\ell_{13}} &
-c_{12}^\ell s_{23}^\ell e^{i\delta^\ell_{23}}-s_{12}^\ell c_{23}^\ell s_{13}^\ell
e^{i(\delta^\ell_{13}-\delta^\ell_{12})} & c_{23}^\ell c_{13}^\ell\\
\end{pmatrix}
\ee
As in the case of the CKM matrix 
the angles $\theta_{ij}^\ell$ can all be made to lie in the first quadrant 
with $0\le \delta^\ell_{12}, \delta^\ell_{23}, \delta^\ell_{13} < 2\pi$.
The matrix $V_{H\nu}$ is then determined through 
$V_{H\nu}=V_{H\ell}V_\text{PMNS}$.

\subsection{The Parameters of the LHT Model}
\label{subsec:2.6}

The new parameters of the LHT model, relevant for the present study, are
\begin{equation}\label{2.16}
f\,,\quad  m^\ell_{H1}\,,\quad m^\ell_{H2}\,,\quad m^\ell_{H3}\,,\quad \theta_{12}^\ell\,,\quad \theta_{13}^\ell\,,\quad \theta_{23}^\ell\,,\quad \delta_{12}^\ell\,\quad \delta_{13}^\ell\,\quad \delta_{23}^\ell\,,
\end{equation}
and the ones in the mirror quark sector that can be probed by FCNC processes in $K$ and $B$ meson systems, as discussed in detail in  \cite{BBPTUW,Blanke:2006eb}.

The determination of  the parameters in \eqref{2.16} with the help of LFV
processes is clearly a formidable task. 
However, if the new particles present in the LHT model are discovered
  once LHC  starts its operation, the parameter $f$ will be determined from
  $M_{W_H}$, $M_{Z_H}$ or $M_{A_H}$. Similarly the mirror lepton masses $m^\ell_{Hi}$ will be measured.

The only remaining free parameters among the ones listed in
\eqref{2.16} will then be $\theta_{ij}^\ell$ and $\delta_{ij}^\ell$, which can be determined once many LFV processes have been measured.

\boldmath
\newsection{$\ell_i\to\ell_j\gamma$ in the LHT Model}\label{sec:liljgamma}
\unboldmath

\subsection{Preliminaries}

In \cite{BBPTUW} we have shown how one can obtain the branching ratio $Br(B\to X_s\gamma)$ in the LHT model directly from the $B\to X_s\gamma$ and $b\to s\,\text{gluon}$ decays in the SM by simply changing the arguments of the two SM functions $D'_0(x)$ and $E'_0(x)$ and adjusting properly various overall factors. The explicit formulae for these functions are given in Appendix \ref{sec:app2}.

Here we will proceed in an analogous way. We will first derive $Br(\mu\to e\gamma)$ in the SM for arbitrary neutrino masses from the calculation of $Br(B\to X_s\gamma)$ in the same model. This will allow us to obtain in a straightforward manner $Br(\mu\to e\gamma)$ in the LHT model, when also some elements of the $Br(B\to X_s\gamma)$ calculation in this model in  \cite{BBPTUW}  are taken into account. The generalization to $\tau\to\mu\gamma$ and $\tau\to e\gamma$ will be automatic.

The current experimental upper bounds for $\mu\to e\gamma$, $\tau\to\mu\gamma$ and $\tau\to e\gamma$ are given by  \cite{muegamma,Banerjee}\footnote{The bounds in \cite{Banerjee} have been obtained by combining Belle \cite{Belle-radiative} and BaBar \cite{BaBar-radiative} results.}
\begin{gather}
Br(\mu\to e\gamma)<1.2\cdot 10^{-11}\,,\\
{Br(\tau\to \mu\gamma)<1.6\cdot 10^{-8}\,,\qquad
Br(\tau\to e\gamma)<9.4\cdot 10^{-8}\,.}
\end{gather}

\boldmath
\subsection{$\mu\to e\gamma$ in the SM}
\unboldmath

\begin{figure}
\center{\epsfig{file=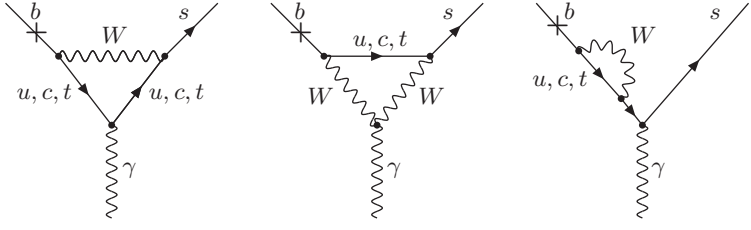}}
\caption{\it Diagrams contributing to $B\to X_s\gamma$ in the SM.\label{fig:BsgSM}}
\end{figure}

\begin{figure}
\center{\epsfig{file=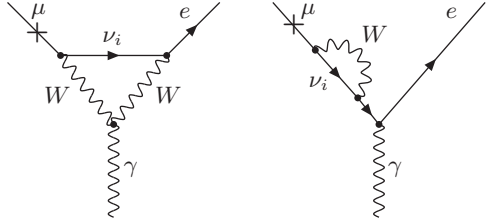}}
\caption{\it Diagrams contributing to $\mu\to e\gamma$ in the SM.\label{fig:megSM}}
\end{figure}

The diagrams for the $B\to X_s\gamma$ and $\mu\to e\gamma$ decays in the SM
are shown in Figs.~\ref{fig:BsgSM} and \ref{fig:megSM}, respectively. In
$R_\xi$ gauge also the corresponding diagrams with Goldstone bosons
contribute. The diagram with {the photon} coupled directly to the internal fermion line, present in $B\to X_s\gamma$, is absent in the case of $\mu\to e\gamma$ due to the neutrino charge neutrality.

In the case of the $B\to X_s\gamma$ decay the function $D'_0(x)$ resulting from the diagrams in Fig.~\ref{fig:BsgSM} can be decomposed as follows
\be\label{eq:Dprime}
D'_0(x) =\left[D'_0(x)\right]_\text{Abelian} + \left[D'_0(x)\right]_\text{triple}\,,
\ee
where the first term on the r.\,h.\,s. represents the sum of the first and the last diagram in Fig.~\ref{fig:BsgSM} and the second term the second diagram.

The inspection of an explicit calculation of the $b\to s\gamma$ transition in the 't~Hooft-Feynman gauge gives
\be\label{eq:Dabelian}
\left[D'_0(x)\right]_\text{Abelian}=(2 Q_d - Q_u)E'_0(x)\,,
\ee
with $Q_d$ and $Q_u$ being the electric charges of external and internal fermions, respectively. Setting $Q_d=-1/3$ and {$Q_u=2/3$} in \eqref{eq:Dabelian} and using \eqref{eq:Dprime} we find
\be\label{eq:Dtriple}
\left[D'_0(x)\right]_\text{triple}= D'_0(x) +\frac{4}{3}E'_0(x)\,.
\ee
This contribution is independent of fermion electric charges and can be directly used in the $\mu\to e\gamma$ decay.

Turning now our attention to the latter decay we find first from \eqref{eq:Dabelian}
\be\label{eq:Dabmeg}
 \left[D'_0(x)\right]_\text{Abelian}^{\mu\to e\gamma}=-2 E'_0(x)\,,
\ee
as $Q_u=0$ and $Q_d=-1$ in this case. The final result for the relevant short distance function in the case of  $\mu\to e\gamma$ is then given by the sum of \eqref{eq:Dtriple} and \eqref{eq:Dabmeg}. Denoting this function by $H(x_\nu^i)$ we find
\be\label{eq:G}
H(x_\nu^i)=D'_0(x_\nu^i) -\frac{2}{3}E'_0(x_\nu^i)\,,\qquad x_\nu^i=\left(\frac{m^i_\nu}{M_W}\right)^2\,.
\ee

The generalization of the known SM result to arbitrary neutrino masses reads then
\be\label{eq:BrmegSM}
Br(\mu\to e\gamma)_\text{SM}=\frac{3\alpha}{2\pi}
\left|\sum_i V^{}_{ei}V^{*}_{\mu i}H(x_\nu^i)\right|^2\,,
\ee
with $V_{ij}$ being the elements of the PMNS matrix.

Now, in the limit of small neutrino masses,
\be
H(x_\nu^i){\longrightarrow} \frac{x^i_\nu}{4}\qquad \text{as}\quad{x^i_\nu\rightarrow0}\,,
\ee
and we confirm the known result
\be
Br(\mu \to e\gamma)_\text{SM}=\frac{3\alpha}{32\pi}
\left|\sum_i V^{}_{ei}V^{*}_{\mu i}x_\nu^i\right|^2\,.
\ee

Assuming that {the angle $\theta_{13}$ of the PMNS matrix} is not very small so that $\Delta m_\text{atm}^2$ dominates the branching ratio, we find
\be
Br(\mu \to e\gamma)_\text{SM} \simeq 0.015 \,\alpha\, s_{13}^2 \left[\frac{\Delta m_\text{atm}^2}{M_W^2}\right]^2 < 10^{-54}\,,
\ee
 where $\Delta m^2_\text{atm} \simle 3.2\cdot 10^{-3}\, \text{eV}^2$ and $s_{13}<0.2$ \cite{PMNSparameters}.

\boldmath
\subsection{$\mu\to e\gamma$ in the LHT Model}
\unboldmath

\begin{figure}
\center{\epsfig{file=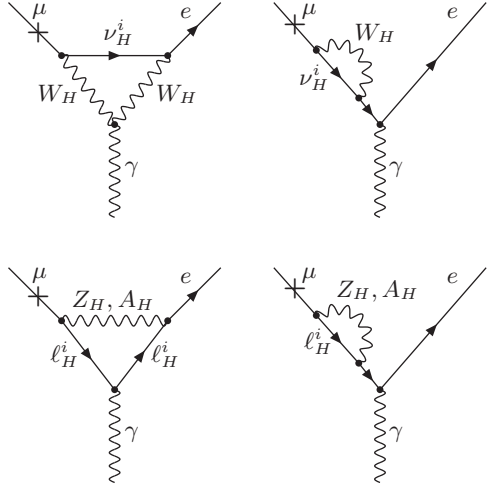}}
\caption{\it Diagrams contributing to $\mu\to e\gamma$ in the LHT model.\label{fig:megLHT}}
\end{figure}

The diagrams contributing to $\mu\to e\gamma$ in the LHT model are shown in Fig.~\ref{fig:megLHT}. We show only the contributions from the mirror fermion sector as the T-even sector gives a negligible contribution. Note that the heavy scalar triplet $\Phi$ does not contribute at this order in $v/f$ (see \cite{BBPTUW,Blanke:2006eb} for details).

Let us write the resulting branching ratio as follows
\be\label{eq:BrmegLHT}
 Br(\mu \to e\gamma)_\text{LHT}=\frac{3\alpha}{2\pi}
\left|\Delta_{W_H}+\Delta_{Z_H}+\Delta_{A_H}\right|^2\,,
\ee
with the different terms representing the $W_H^\pm$, $Z_H$ and $A_H$ contributions.

Defining
\be\label{eq:yi}
y_i=\frac{{m^\ell_{Hi}}^2}{M_{W_H}^2}\,,\qquad y_{i}^{\prime}= a \, y_{i} \quad\text{with }\; a = \frac{5}{\tan^{2} \theta_{W}}\simeq 16.6\,,
\ee
and including the factor
\be
\frac{M_{W_L}^2}{M_{W_H}^2}=\frac{1}{4}\frac{v^2}{f^2}\,,
\ee
we find, using \eqref{eq:BrmegSM},
\be\label{eq:DWH}
\Delta_{W_H}=\frac{1}{4}\frac{v^2}{f^2}\sum_i\chi_i^{(\mu e)}H(y_i)\,,
\ee
with $H$ defined in \eqref{eq:G} and $\chi_i^{(\mu e)}$ in \eqref{eq:chi}.

The neutral gauge boson contributions can directly be deduced from (4.10) of \cite{BBPTUW}. Including a 
{factor 3}
that takes into account the difference between the electric charges of quarks and leptons, we obtain from the last two terms in (4.10) of \cite{BBPTUW}
\bea
\Delta_{Z_H}&=&\frac{1}{4}\frac{v^2}{f^2}\sum_i\chi_i^{(\mu e)}\left[-\frac{1}{2}E'_0(y_i)\right],\label{eq:DZH}\\
\Delta_{A_H}&=&\frac{1}{4}\frac{v^2}{f^2}\sum_i\chi_i^{(\mu e)}\left[-\frac{1}{10}E'_0(y^{\prime}_i)\right].\label{eq:DAH}
\eea

Finally, adding the three contributions in \eqref{eq:DWH}--\eqref{eq:DAH}, we find using \eqref{eq:BrmegLHT}
\be
\label{eq:Brmeg}
 Br(\mu \to e\gamma)_\text{LHT}=\frac{3\alpha}{2\pi}\left|\bar D^{\prime\,\mu e}_\text{odd} \right|^2\,,
\ee
with
\be\label{eq:Dpoddme}
\bar D^{\prime\,\mu e}_\text{odd}=\frac{1}{4}\frac{v^2}{f^2}\left[\sum_i\chi_i^{(\mu e)}\left(D'_0(y_i)-\frac{7}{6}E'_0(y_i)-\frac{1}{10}E'_0(y'_i)\right)\right]\,,
\ee
with $\chi_i^{(\mu e)}$ defined in \eqref{eq:chi}, $y_i$ in \eqref{eq:yi} and
$D'_0$, $E'_0$ given in Appendix \ref{sec:app2}. {The formulae} \eqref{eq:Brmeg} and \eqref{eq:Dpoddme} represent the main result of this section.

 Let us next compare our result with the one of Goyal \cite{Goyal}, which has
been obtained without including the interferences between $\Delta_{W_H}$,
$\Delta_{Z_H}$ and $\Delta_{A_H}$, whose omission cannot be justified. His results for  $|\Delta_{W_H}|^2$, $|\Delta_{Z_H}|^2$ and
$|\Delta_{A_H}|^2$ agree with ours, provided the latter contribution is
divided by 25 in (15) of
\cite{Goyal}, 
once one assumes the straightforward relation $F_{A_H}(Z_{A_H})=F_{Z_H}(a Z_{Z_H})$
between the $A_H$ and $Z_H$ short distance functions.  
Note that the short distance functions $F_j(z_i)$ in \cite{Goyal} still contain mass independent terms that disappear in the final expressions after the unitarity of the $V_{H\ell}$ matrix has been used. For this reason we have removed such terms from the functions $D'_0(x)$ and $E'_0(x)$. Finally, in contrast to \cite{Goyal}, we find that the contributions of the scalar triplet $\Phi$ are of higher order in $v/f$. 

In a recent paper by Choudhury {\it et al.} \cite{IndianLFV} a new analysis of
$\ell_i\to\ell_j\gamma$ has been presented in which the interference terms
have been taken into account. Unfortunately, the formulae in that paper are
{quite complicated thus preventing us from making an analytic
  comparison. On the other hand, we have performed a numerical comparison,
  finding significant differences between our results and those in \cite{IndianLFV}.}

\boldmath
\subsection{$\tau\to e\gamma$ and $\tau\to \mu\gamma$  in the LHT Model}
\unboldmath

The branching ratios for $\tau\to e\gamma$ and $\tau\to\mu\gamma$ can easily be found in analogy to $\mu\to e\gamma$. We find
\bea
Br(\tau\to e\gamma)&=&\frac{3\alpha}{2\pi}
Br(\tau^-\to\nu_\tau e^-\bar\nu_e)\left|\bar D^{\prime\,\tau e}_\text{odd}\right|^2\,,\\
Br(\tau\to \mu\gamma)&=&\frac{3\alpha}{2\pi}
Br(\tau^-\to\nu_\tau \mu^-\bar\nu_\mu)\left|\bar D^{\prime\,\tau \mu}_\text{odd}\right|^2\,,
\eea
where $\bar D^{\prime\,\tau e}_\text{odd}$ and $\bar D^{\prime\,\tau \mu}_\text{odd}$ can be obtained from \eqref{eq:Dpoddme} by replacing $(\mu e)$ with $(\tau e)$ and $(\tau\mu)$, respectively. Furthermore \cite{PDG}
\be
Br(\tau^-\to\nu_\tau e^-\bar\nu_e)=(17.84\pm0.05)\%\,,\qquad Br(\tau^-\to\nu_\tau \mu^-\bar\nu_\mu)=(17.36\pm0.05)\%\,.
\ee

\newsection{Semi-leptonic $\bm{\tau}$ Decays}\label{sec:semilep}

{Recently the  Belle \cite{Belle-semi} and BaBar \cite{BaBar-semi} collaborations presented improved upper bounds for the decays $\tau\to\ell P$ $(P=\pi,\eta,\eta')$, which have been combined \cite{Banerjee} to}
\begin{gather}
{Br(\tau\to\mu\pi)<5.8\cdot10^{-8}\,,\qquad
Br(\tau\to\mu\eta)<5.1\cdot10^{-8}\,,\qquad
Br(\tau\to\mu\eta')<5.3\cdot10^{-8}\,,}\\
{
Br(\tau\to e\pi)<4.4\cdot10^{-8}\,,\qquad
Br(\tau\to e\eta)<4.5\cdot10^{-8}\,,\qquad
Br(\tau\to e\eta')<9.0\cdot10^{-8}\,,}
\end{gather}
thus increasing the interest in investigating these branching ratios in the LHT model.

The branching ratios for these semi-leptonic decays have already been
estimated within the LHT model by Goyal \cite{Goyal}, but with the rough
approximation consisting in the use of the effective Hamiltonian for $K^0-\bar K^0$ mixing.
Here we will study the semileptonic decays in question with a more
accurate approach, having at hand the recent analysis of rare $K$ and $B$ decays in the LHT model \cite{Blanke:2006eb}.

\begin{figure}
\center{\epsfig{file=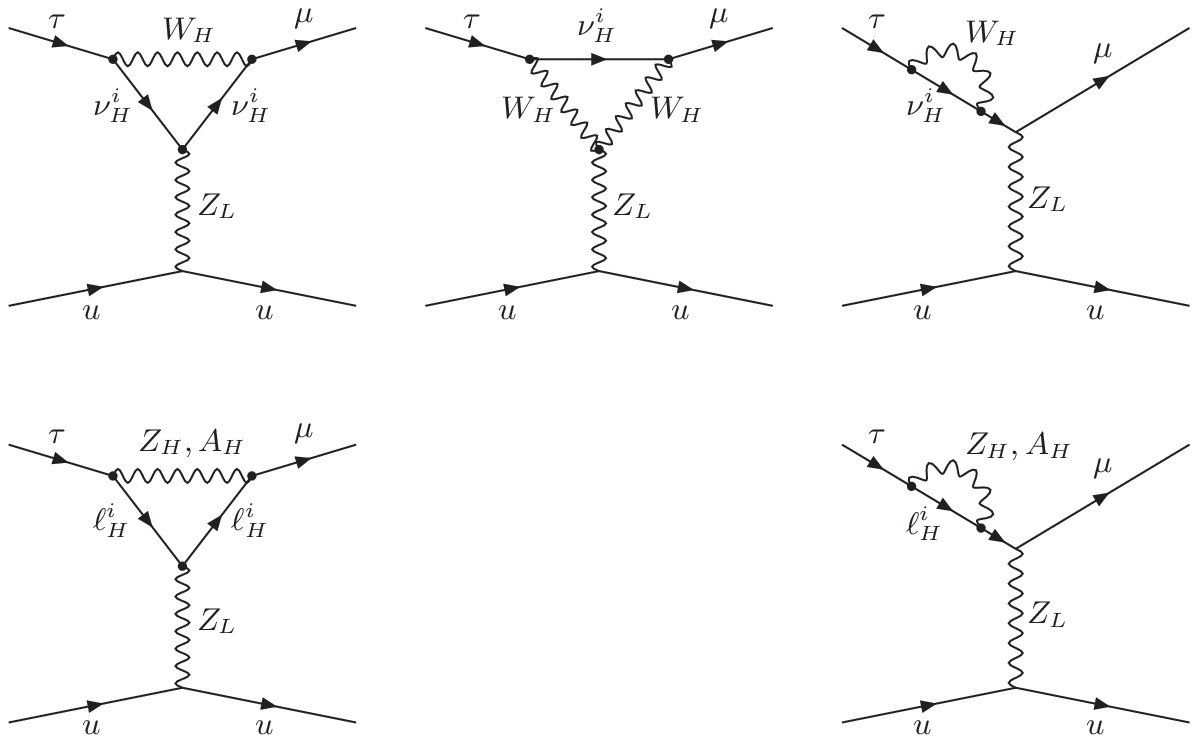}}
\vspace{0.5cm}

\center{\epsfig{file=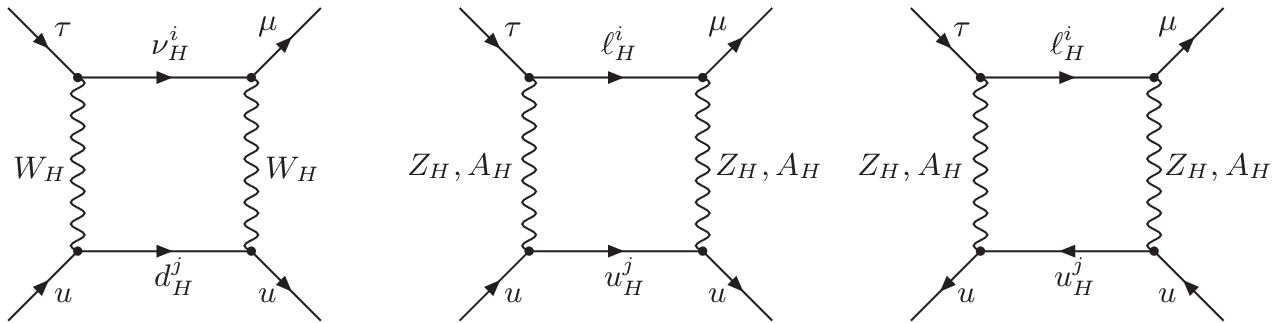}}
\caption{\it Diagrams contributing to $\tau\to\mu\pi$ in the LHT model.\label{fig:taumupi} Similar diagrams, but with $d$ quarks in the final state, also contribute.}
\end{figure}

The diagrams for $\tau\to\mu\pi$ are shown in Fig.~\ref{fig:taumupi}. As $\pi^0$ has the following flavour structure
\be\label{eq:pi0}
\pi^0=\frac{\bar uu-\bar dd}{\sqrt{2}}\,,
\ee
there are two sets of diagrams, with $\bar uu$ and $\bar dd$ in the final
state. The corresponding effective Hamiltonians can directly be obtained from
\cite{Blanke:2006eb}. They involve the short-distance functions $\bar
X_\text{odd}^{\tau \mu}$
and $-\bar Y_\text{odd}^{\tau \mu}$ for  $\bar uu$ and $\bar dd$, respectively. Taking
into account the opposite sign that is conventionally chosen to define the
two short distance functions, the effective Hamiltonian that includes both sets of diagrams is simply given as follows
\be
\Heff=\frac{G_F}{\sqrt{2}}\frac{\alpha}{2\pi\sin^2\theta_W}\left(\bar
  X_\text{odd}^{\tau \mu} (\bar uu)_{V-A} - \bar Y_\text{odd}^{\tau \mu} (\bar dd)_{V-A} \right)(\bar\mu\tau)_{V-A}\,.
\ee
Here $\bar X_\text{odd}^{\tau \mu}$ and $\bar Y_\text{odd}^{\tau \mu}$ have
the same structure as the functions calculated in \cite{Blanke:2006eb} in the context of rare $K$ and $B$ decays. Adapting them to the lepton sector we find:
\begin{eqnarray}
\bar{X}_\text{odd}^{\tau \mu}&=& \left[
\chi_2^{(\tau \mu)}\big(J^{u\bar u}(y_2,z)-J^{u\bar u}(y_1,z)\big)
+\chi_3^{(\tau \mu)}\big(J^{u \bar u}(y_3,z)-J^{u \bar u}(y_1,z)\big)
\right],
\label{Xodd}\\
\bar{Y}_\text{odd}^{\tau \mu}&=& \left[
\chi_2^{(\tau \mu)}\big(J^{d\bar d}(y_2,z)-J^{d \bar d}(y_1,z)\big)
+\chi_3^{(\tau \mu)}\big(J^{d\bar d}(y_3,z)-J^{d\bar d}(y_1,z)\big)
\right] ,\label{Yodd}
\end{eqnarray}
where
\begin{eqnarray}
J^{u \bar{u}}\left(y_{i}, z\right)&=&
\frac{1}{64}\frac{v^2}{f^2}\bigg[y_i S_\text{odd} +F^{u
    \bar{u}}(y_i,z;W_H)\nn\\
&& \qquad +4
\Big(G(y_i,z;Z_H)+G_1(y'_i,z';A_H)+G_2(y_i,z;\eta)\Big)\bigg],
\label{Znunu}\\
J^{d\bar d}\left(y_{i}, z\right)&=&
\frac{1}{64}\frac{v^2}{f^2}\bigg[{y_i} S_\text{odd} +F^{d\bar d}(y_i,z;W_H)\nn\\
&& \qquad -4
\Big(G(y_i,z;Z_H)+G_1(y'_i,z';A_H)-G_2(y_i,z;\eta)\Big)\bigg]
,\label{Zmumu}\\
S_\text{odd}&=&\frac{1}{\varepsilon}+\log\frac{\mu^2}{M_{W_H}^2} \longrightarrow \log\frac{(4\pi f)^2}{M_{W_H}^2}\label{eq:Sodd}\,,
\end{eqnarray}
with the functions $F^{u \bar{u}}$, $F^{d \bar{d}}$, $G$, $G_{1}$ and $G_{2}$ given in Appendix
\ref{sec:app2}, the leptonic variables $y_i$ and $y_i^{\prime}$ defined in
(\ref{eq:yi}) and the analogous variables for degenerate mirror 
quarks\footnote{The limit of 
  degenerate mirror quarks represents here a good approximation, as the box
  contributions vanish in the limit of degenerate mirror leptons and,
  consequently, the inclusion of mass splittings in the quark sector represents
  a higher order effect.} given by
\be
z = \frac{{m^q_{H}}^2}{M_{W_H}^2}\,, \qquad z^{\prime}= a\, z\,,
\quad \eta = \frac{1}{a}\,.
\ee

As
\be
\langle 0|(\bar uu)_{V-A}|\pi^0\rangle=-\langle 0|(\bar dd)_{V-A}|\pi^0\rangle = \frac{F_\pi p_\pi^\mu}{\sqrt{2}}\,,
\ee
where $F_\pi \simeq 131 \mev$ is the pion decay constant, we find
\be\label{eq:Brtaumupi}
Br(\tau\to\mu\pi)=\frac{G_F^2 \alpha^2 F_\pi^2 m_\tau^3 \tau_\tau}{128\pi^3 \sin^4
  \theta_W }\, |\bar X_\text{odd}^{\tau \mu} + \bar Y_\text{odd}^{\tau \mu}|^2\,,
\ee
with $\tau_\tau$ and $m_\tau$ being the lifetime and mass of the decaying
$\tau$, and neglecting suppressed pion and muon mass contributions of the order
$\mathcal{O}(m_\pi^2/m_\tau^2)$ and $\mathcal{O}(m_\mu^2/m_\tau^2)$. The branching ratio for the $\tau\to e\pi$ decay can  be obtained very easily from \eqref{eq:Brtaumupi} by simply replacing $(\tau\mu)$ with $(\tau e)$. 
The generalization of \eqref{eq:Brtaumupi} to the decays $\tau\to\mu\eta$ and
$\tau\to\mu \eta'$ is quite straightforward too, although slightly complicated
by mixing in the $\eta - \eta'$ system.

The understanding of $\eta - \eta'$ mixing has largely improved in the last
decade mainly thanks to the formulation of a new mixing scheme~\cite{Kaiser:1998ds,Feldmann:1997vc}, where
not one but two angles are introduced to relate the physical states ($\eta$,
$\eta'$) to the octet and singlet states ($\eta_8$, $\eta_0$), as
\be
| \eta \rangle =\cos \theta_8 | \eta_8 \rangle - \sin \theta_0 | \eta_0
\rangle\,,\qquad
| \eta' \rangle =\sin \theta_8 | \eta_8 \rangle + \cos \theta_0 | \eta_0
\rangle\,,
\ee
with
\be
| \eta_8 \rangle =\frac{1}{\sqrt{6}}(| \bar u u \rangle + | \bar d d \rangle -
2 | \bar s s \rangle)\,,\qquad
| \eta_0 \rangle =\frac{1}{\sqrt{3}}(| \bar u u \rangle + | \bar d d \rangle + | \bar s s \rangle)\,.
\ee 
In this mixing scheme, four independent decay constants are involved.
Each of the two physical mesons ($P=\eta, \eta'$), in fact, has both octet
($a=8$) and singlet ($a=0$) components, defined by
\be
\langle 0 | (\bar q {\frac{\lambda^a}{2}} q)_{V-A} | P(p) \rangle =
\frac{F_P^a p_\mu}{\sqrt{2}}\,,
\ee
where the $SU(3)$ generators $\lambda^a$ satisfy the normalization convention
$\text{Tr}[ \lambda^a \lambda^b] = 2 \delta^{a b}$.
They are conveniently parameterized~\cite{Kaiser:1998ds} in terms of the two mixing angles
($\theta_8, \theta_0$) and two basic decay constants ($F_8, F_0$), as
\be
\begin{pmatrix} F_\eta^8 & F_\eta^0\\  F_{\eta'}^8 & F_{\eta'}^0\\
  \end{pmatrix}\, = \begin{pmatrix} F_8 \cos \theta_8 & -F_0 \sin
    \theta_0\\  F_8 \sin \theta_8 & F_0 \cos \theta_0\\
  \end{pmatrix}\,.
\ee 

Working in this mixing scheme and generalizing the expression for the $\tau
\to \mu \pi$ branching ratio in (\ref{eq:Brtaumupi}), one obtains
\bea\label{eq:Brtaumueta}
Br(\tau\to\mu\eta)&=&\frac{G_F^2 \alpha^2 F_\pi^2 m_\tau^3 \tau_\tau}{128\pi^3 \sin^4
  \theta_W }\, \left|\frac{\cos \theta_8}{\sqrt{3}}\frac{F_8}{F_\pi}(\bar X_\text{odd}^{\tau \mu} + \bar Y_\text{odd}^{\tau \mu})
- \sqrt{\frac{2}{3}}\sin \theta_0\frac{F_0}{F_\pi}(\bar X_\text{odd}^{\tau
  \mu}  -2\, \bar Y_\text{odd}^{\tau \mu})\right|^2\,,\nn\\
Br(\tau\to\mu\eta')&=&\frac{G_F^2 \alpha^2 F_\pi^2 m_\tau^3 \tau_\tau}{128\pi^3 \sin^4
  \theta_W }\, \left| \frac{\sin \theta_8}{\sqrt{3}}\frac{F_8}{F_\pi}(\bar X_\text{odd}^{\tau \mu} + \bar Y_\text{odd}^{\tau \mu})
+ \sqrt{\frac{2}{3}}\cos \theta_0\frac{F_0}{F_\pi}(\bar X_\text{odd}^{\tau
  \mu}  -2\, \bar Y_\text{odd}^{\tau \mu})\right|^2\,.\nn\\
\eea

We conclude this section noting  that the function $S_\text{odd}$ in (\ref{eq:Sodd}) represents a
left-over singularity that signals some sensitivity
  of the final results to the UV completion of the theory. 
This issue is known from the study of
  electroweak precision constraints \cite{tparity} and has been discussed in detail in the context of the LH model
  without T-parity \cite{BPUB} and, more recently, also with T-parity
  \cite{Blanke:2006eb}. We refer to the latter paper for a detailed discussion.
Here we just mention that, in estimating the contribution of these logarithmic
singularities as in \eqref{eq:Sodd},
 we have  assumed the UV completion of the theory not to have a 
{complicated}
  flavour pattern or at least that it has no impact below the cut-off.
Clearly, this
additional
 assumption lowers the predictive power of the theory. 
In spite of that, we believe
 that the general picture of 
{LFV} 
processes presented here
is only insignificantly shadowed by this general property of
 non-linear sigma models.

\boldmath
\newsection{$\mu^-\to e^-e^+e^-$, $\tau^-\to\mu^-\mu^+\mu^-$ and $\tau^-\to e^-e^+e^-$}\label{sec:mueee}
\unboldmath

Next, we will consider the decay
$\mu^-\to e^-e^+e^-$, for which the experimental upper bound reads \cite{meee}
\be
Br(\mu^-\to e^-e^+e^-)< 1.0 \cdot10^{-12}\,.
\ee
This decay is governed, analogously to the $b\to s\mu^+\mu^-$ transition,
analyzed in the LHT model already in \cite{Blanke:2006eb}, by contributions from
$\gamma$- and $Z^0$-penguins and by box diagrams. However, the fact that now
in the final state two identical particles are present does not allow to use
directly the known final expressions for $b\to s\mu^+\mu^-$, although some
intermediate results from the latter decay turned out to be useful here. Also
the general result for $\mu^-\to e^-e^+e^-$ {obtained in \cite{Hisano}}, which has been corrected in \cite{Ellis,Herrero}, turned out to be very helpful.

Performing the calculation in the unitary gauge, where we found the contribution from the $Z^0$-penguin to vanish \cite{Blanke:2006eb}, we find for the relevant amplitudes from  photon penguins and box diagrams\footnote{Following \cite{Hisano}, our sign conventions are chosen such that $\Heff$ is determined from $-\mathcal{A}$.}:
\bea\label{eq:Agamma'}
\mathcal{A}_{\gamma'} &=& \frac{G_F}{\sqrt{2}}\frac{e^2}{8\pi^2}\frac{1}{q^2}\bar D^{\prime\,\mu e}_\text{odd} \left[\bar e(p_1)(m_\mu i\sigma_{\alpha\beta} q^\beta (1+\gamma_5))\mu(p) \right]\otimes \left[\bar e(p_2)\gamma^\alpha e(p_3)\right]\nn\\
&&\hspace{2cm} -(p_1 \leftrightarrow p_2)\,,\qquad\\
\mathcal{A}_{\gamma} &=& -\left[4 \frac{G_F}{\sqrt{2}} \frac{e^2}{8\pi^2} \bar Z^{\mu e}_\text{odd}\left[\bar e(p_1)\gamma_\alpha(1-\gamma_5)\mu(p)\right]\otimes\left[\bar e(p_2)\gamma^\alpha e(p_3)\right] -(p_1 \leftrightarrow p_2)\right]\,,\qquad\\
\mathcal{A}_\text{box} &=& 2 \frac{G_F}{\sqrt{2}}\frac{\alpha}{2\pi\sin^2\theta_W}\bar Y^{\mu e}_{e,\text{odd}} \left[\bar e(p_1)\gamma_\alpha(1-\gamma_5)\mu(p)\right]\otimes\left[\bar e(p_2)\gamma^\alpha(1-\gamma_5) e(p_3)\right] \,.
\eea 
The function $\bar D^{\prime\,\mu e}_\text{odd}$ is given in \eqref{eq:Dpoddme}, while the functions $\bar Y_{e,\text{odd}}^{\mu e}$ and $\bar Z_\text{odd}^{\mu e}$ can easily be obtained from those calculated in \cite{Blanke:2006eb}.
The analogy with the $b\to s\mu^+\mu^-$ decay, together with the observation
that the $\mu^-\to e^-e^+e^-$ decay in question involves only leptons in both
the initial and final states, allow us to write\footnote{The subscript $e$ of $\bar Y_{e,\text{odd}}^{\mu e}$ denotes which of the SM charged leptons appears on the flavour conserving side of the relevant box diagrams.}
\bea
 \bar Y_{e,\text{odd}}^{\mu e}&=&\chi_2^{(\mu e)} \sum_{i=1}^{3} |V_{H \ell}^{i
   e}|^2 \left[ J^{d \bar d}(y_2,y_i) - J^{d \bar d}(y_1,y_i)\right]\nn\\
&& +\,
\chi_3^{(\mu e)} \sum_{i=1}^{3} |V_{H \ell}^{i
   e}|^2 \left[ J^{d \bar d}(y_3,y_i) - J^{d \bar d}(y_1,y_i)\right] \label{eq:Yoddme}
\eea
with $J^{d\bar d}$ given in \eqref{Zmumu}.
Following a similar reasoning we can write for the $\bar Z_\text{odd}^{\mu e}$ function
\be\label{eq:Zoddme}
\bar Z_{\rm odd}^{\mu e}=\bigg[ \chi^{(\mu e)}_2 \big(Z_{\rm odd}(y_2)-Z_{\rm
    odd}(y_1)\big)+ \chi^{(\mu e)}_3 \big(Z_{\rm odd}(y_3)-Z_{\rm
    odd}(y_1)\big) \bigg] \,,
\ee
where 
\be
Z_{\rm odd}(y_i)=C_{\rm odd}(y_i)+\frac{1}{4} D_{\rm odd}(y_i)\,.
\label{eq:ZCDodd}
\ee

The explicit expressions of the $C_\text{odd}$ and $D_\text{odd}$ functions are
given in Appendix \ref{sec:app2}\footnote{Note that the functions $C_\text{odd}$ and $D_\text{odd}$ are gauge dependent and have been calculated in the 't~Hooft-Feynman gauge. However, the function $\bar Z_{\rm odd}^{\mu e}$ is gauge independent, so that it can be used also in the  unitary gauge calculation above.}. Here, we just note that as a consequence of
the charge difference between the leptons involved in $\mu^-\to e^-e^+e^-$ and
the quarks involved in $b\to s\mu^+\mu^-$, $D_{\rm odd}$ in (\ref{eq:ZCDodd}) differs from the analogous function found in  \cite{Blanke:2006eb}.

Comparing these expressions to the general expressions for the amplitudes given in \cite{Hisano,Herrero}, we easily obtain $\Gamma(\mu^-\to e^-e^+e^-)$. Normalizing by $\Gamma(\mu^-\to e^-\bar\nu_e\nu_\mu)$, we find the branching ratio for the decay $\mu^-\to e^-e^+e^-$ to be 
\bea
Br(\mu^-\to e^-e^+e^-) &=& \frac{\Gamma(\mu^-\to e^-e^+e^-)}{\Gamma(\mu^-\to e^-\bar\nu_e\nu_\mu)} \nn\\
&=& \frac{\alpha^2}{\pi^2}\bigg[ 3\left| \bar Z^{\mu e}_\text{odd} \right|^2 + 3\,\text{Re}\left(\bar Z^{\mu e}_\text{odd} (\bar D^{\prime\,\mu e}_\text{odd})^*\right)+\left|\bar D^{\prime\,\mu e}_\text{odd}\right|^2\left(\log\frac{m_\mu}{m_e}-\frac{11}{8}\right)\nn\\
&&\qquad +\frac{1}{2\sin^4\theta_W}\left| \bar Y^{\mu e}_{e,\text{odd}} \right|^2-\frac{2}{\sin^2\theta_W}\text{Re}\left( \bar Z^{\mu e}_\text{odd} (\bar Y^{\mu e}_{e,\text{odd}})^* \right)\nn\\
&&\qquad-\frac{1}{\sin^2\theta_W}\text{Re}\left( \bar D^{\prime\,\mu e}_\text{odd} (\bar Y^{\mu e}_{e,\text{odd}})^* \right)\bigg]\,.\label{eq:Brmeee}
\eea

For $\tau^-\to\mu^-\mu^+\mu^-$ we make the following replacements in \eqref{eq:Agamma'}--\eqref{eq:Brmeee}:
\be
V_{H\ell}^{ie} \to V_{H\ell}^{i\mu}\,,\qquad (\mu e)\to(\tau\mu)\,,\qquad m_\mu \to m_\tau\,,\qquad m_e\to m_\mu\,,
\ee
so that, in particular, $\bar Y^{\tau\mu}_{\mu,\text{odd}}$ is now present. Furthermore, in \eqref{eq:Brmeee} the normalization $\Gamma(\mu^-\to e^-\nu_\mu\bar\nu_e)$ is replaced by $\Gamma(\tau^-\to \mu^-\nu_\tau\bar\nu_\mu)$, so that the final result for $Br(\tau^-\to\mu^-\mu^+\mu^-)$ contains an additional factor $Br(\tau^-\to \mu^-\nu_\tau\bar\nu_\mu)$. In the case of $\tau^-\to e^-e^+e^-$ the replacements in \eqref{eq:Yoddme}--\eqref{eq:Brmeee} amount only to
\be
(\mu e)\to (\tau e)\,,\qquad m_\mu \to m_\tau\,,
\ee
having now $\bar Y^{\tau e}_{e,\text{odd}}$, and in \eqref{eq:Brmeee} $\Gamma(\mu^-\to e^-\nu_\mu\bar\nu_e)$ is replaced by $\Gamma(\tau^-\to e^-\nu_\tau\bar\nu_e)$ and an additional factor $Br(\tau^-\to e^-\nu_\tau\bar\nu_e)$ appears. In doing this we neglect in all three expressions $m_{e,\mu}$ with respect to $m_\tau$.

\newsection{$\bm{\mu-e}$ Conversion in Nuclei}
\label{sec:mueconv}

Similarly to the decays $\mu\to e\gamma$ and $\mu^-\to e^-e^+e^-$, stringent experimental upper bounds on $\mu-e$ conversion in nuclei exist. In particular, the experimental upper bound on $\mu-e$ conversion in $^{48}_{22}\text{Ti}$ reads \cite{mue-conv_bound} 
\be
R(\mu\text{Ti}\to e\text{Ti})<4.3\cdot 10^{-12}\,,
\ee
{and the dedicated J-PARC experiment PRISM/PRIME  should reach a sensitivity of $\ord(10^{-18})$ \cite{J-PARK}.}

A very detailed calculation of the $\mu-e$ conversion rate in various nuclei has been performed in \cite{Kitano}, using the methods developed by Czarnecki {\it et al.} \cite{Czarnecki}. {It has been emphasized in \cite{Kitano} that the atomic number dependence of the conversion rate can be used to distinguish between different theoretical models of LFV. Useful general formulae can also be found in \cite{Hisano}.}

{We have calculated the $\mu-e$ conversion rate in nuclei in the LHT model using the general model-independent formulae of both \cite{Hisano} and \cite{Kitano}. We have checked numerically that, for relatively light nuclei such as Ti, both results agree within 10\%. Therefore, we will give the result for $\mu-e$ conversion in nuclei derived from the general expression given in \cite{Hisano}, as it has a more transparent structure than the one of \cite{Kitano }.} 

Following a similar reasoning as in the previous section, we find from (58) of  \cite{Hisano}
\bea\label{eq:mueconv}
\Gamma(\mu\text{X} \to e\text{X})&=&\frac{G_F^2}{8\pi^4}\alpha^5\frac{Z_\text{eff}^4}{Z}|F(q)|^2m_\mu^5 \nn\\
&&\cdot\left| Z\left(4\bar Z^{\mu e}_\text{odd}+\bar D^{\prime\,\mu e}_\text{odd} \right)-(2Z+N)\frac{\bar X^{\mu e}_\text{odd}}{\sin^2\theta_W}+(Z+2N)\frac{\bar Y^{\mu e}_\text{odd}}{\sin^2\theta_W}  \right|^2\,,\qquad
\eea
where $\bar X^{\mu e}_\text{odd}$ and $\bar Y^{\mu e}_\text{odd}$ are obtained from \eqref{Xodd} and \eqref{Yodd} by making the replacement $(\tau\mu) \to (\mu e)$, and $\bar D^{\prime\,\mu e}_\text{odd}$ and $\bar Z^{\mu e}_\text{odd}$ are given in \eqref{eq:Dpoddme} and \eqref{eq:Zoddme}, respectively. $Z$ and $N$ denote the proton and neutron number of the nucleus. $Z_\text{eff}$ has been determined in \cite{Zeff} and $F(q^2)$ is the nucleon form factor.  For $\text{X}={}^{48}_{22}\text{Ti}$,  $Z_\text{eff}=17.6$ and $F(q^2\simeq-m_\mu^2)\simeq 0.54$ \cite{mueconv_Zprime}.

The $\mu-e$ conversion rate $R(\mu\text{X}\to e\text{X})$ is then given by
\be
R(\mu\text{X}\to e\text{X})=\frac{\Gamma(\mu\text{X} \to e\text{X})}{\Gamma^\text{X}_\text{capture}}\,,
\ee
with $\Gamma^\text{X}_\text{capture}$ being the $\mu$ capture rate of the element X. The experimental value is given by $\Gamma^\text{Ti}_\text{capture}=
(2.590\pm0.012)\cdot10^6\,\text{s}^{-1}$ \cite{capture}.

In our numerical analysis of Section \ref{sec:num} we will restrict ourselves to $\mu-e$ conversion in $^{48}_{22}\text{Ti}$, {for which the most stringent experimental upper bound exists and where the approximations entering \eqref{eq:mueconv}  work very well. For details, we refer the reader to \cite{Hisano,Kitano,mueconv_Zprime}.}

\newsection{$\bm{K_{L,S}\to\mu e}$ and $\bm{K_{L,S}\to\pi^0\mu e}$ in the LHT Model}\label{sec:KLmue}

The rare decay $K_L\to\mu e$ is well known from the studies of the Pati-Salam (PS) model~\cite{PS},
where it proceeds through a tree level leptoquark exchange in the
t-channel. The stringent upper bound on its rate \cite{KLmue-exp}
\be\label{eq:KLmueexp} 
Br(K_L\to\mu e)= Br(K_L\to\mu^+ e^-)+Br(K_L\to\mu^- e^+)<4.7\cdot10^{-12}\,,
\ee
implies the mass of the leptoquark gauge boson to be above $10^3\tev$ \cite{leptoquarks}. By increasing the weak gauge group of the PS model in the context of the so-called ``Petite Unification'' \cite{PU} and placing ordinary fermions in multiplets with heavy new fermions, and not with the ordinary SM fermions as in the PS model, it is possible to avoid tree level contributions to $K_L\to\mu e$ so that the process is dominated by the box diagrams with new heavy gauge bosons, heavy quarks and heavy leptons exchanged. The relevant masses of these particles can then be decreased to $\ord(1\tev)$ without violating the bound in \eqref{eq:KLmueexp}.

In the SM the decay $K_L\to\mu e$  is forbidden at the tree level but can
proceed through box diagrams in the case of non-degenerate neutrino
masses. Similarly to $\mu\to e\gamma$ it is too small to be measured.

\begin{figure}
\center{\epsfig{file=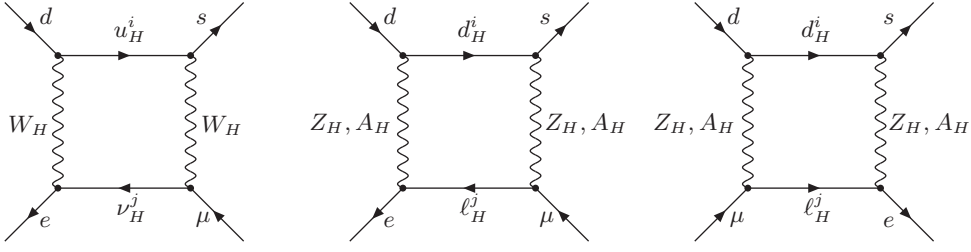}}
\caption{\it Diagrams contributing to $K_L\to\mu^+ e^-$ in the LHT model. Similar diagrams describe $K_L\to\mu^- e^+$.\label{fig:KLmue}}
\end{figure}

{In the LHT model as well, $K_L\to\mu e$ appears first at one loop level.} It proceeds through the diagrams shown in Fig.~\ref{fig:KLmue} that should be compared with very similar diagrams in Fig.~1 of \cite{BBPTUW} contributing to particle-antiparticle mixing in the T-odd sector of the LHT model. The main difference is the appearance of leptons in the lower part of the box diagrams instead of quarks, leading to a different structure of the loop functions once the unitarity of the matrices $V_{Hd}$ and $V_{H\ell}$ has been used.

The effective Hamiltonian for $K_L\to\mu e$ can be directly obtained from (3.11) of \cite{BBPTUW} by removing the QCD factor $\eta_2$, appropriately changing the CKM-like factors and the arguments of the short distance functions and multiplying by 2 to correct for a different combinatorial factor.

Taking into account that both the $\mu^+e^-$ and $\mu^-e^+$ final states are experimentally detected, the relevant effective Hamiltonian reads
\bea
\Heff&=&\frac{G_F^2}{32\pi^2}M_{W_L}^2\frac{v^2}{f^2}\sum_{i,j}\xi^{(K)}_iF_H(z_i,y_j) \left[\chi^{(\mu e)}_j (\bar s d)_{V-A}(\bar e\mu)_{V-A}\right.\nn\\
&&\hspace{5.5cm}+\left. {\chi^{(\mu e)}_j}^* (\bar s d)_{V-A}(\bar \mu e)_{V-A}\right]+h.c.\,,
\eea
where $\chi^{(\mu e)}_i$ and $y_i$ are defined in (\ref{eq:chi}) and (\ref{eq:yi}),
and $\xi_i^{(K)}$ and $z_i$ belong to the quark
sector and are defined as
\be
\xi_i^{(K)}=V^{*is}_{Hd} V^{id}_{Hd}\,, \qquad z_i=\frac{{m^q_{H i}}^2}{M_{W_H}^2}\,.
\ee
The short distance functions read 
\be\label{eq:FH}
F_H(z_i,y_j)=F(z_i,y_j;W_H)+G(z_i,y_j;Z_H)+A_1(z_i,y_j;Z_H)+
A_2(z_i,y_j;Z_H)\,,
\ee
where the different contributions correspond to ``$WW$'', ``$ZZ$'', ``$AA$'' and ``$ZA$'' diagrams, respectively. Explicit expressions for the functions $F$, $G$, $A_1$ and $A_2$ can be found in Appendix \ref{sec:app2}.

Using the unitarity of the $V_{Hd}$ and $V_{H\ell}$ matrices we find
\bea
\Heff &=& \frac{G_F^2}{32\pi^2}M_{W_L}^2\frac{v^2}{f^2}\Bigg\{ \Big[
\chi_2^{(\mu e)}\xi_2^{(K)} R(z_2,z_1,y_1,y_2) +
\chi_3^{(\mu e)}\xi_3^{(K)} R(z_3,z_1,y_1,y_3)\nn\\
&& \qquad+\, \chi_2^{(\mu e)}\xi_3^{(K)} R(z_3,z_1,y_1,y_2) + 
\chi_3^{(\mu e)}\xi_2^{(K)} R(z_2,z_1,y_1,y_3)\Big](\bar s d)_{V-A}(\bar e\mu)_{V-A}\nn\\
&&+\,\Big[
{\chi_2^{(\mu e)}}^*\xi_2^{(K)} R(z_2,z_1,y_1,y_2) +
{\chi_3^{(\mu e)}}^*\xi_3^{(K)} R(z_3,z_1,y_1,y_3)\nn\\
&&\qquad +\, {\chi_2^{(\mu e)}}^*\xi_3^{(K)} R(z_3,z_1,y_1,y_2) + 
{\chi_3^{(\mu e)}}^*\xi_2^{(K)} R(z_2,z_1,y_1,y_3)\Big](\bar s d)_{V-A}(\bar \mu e)_{V-A}\Bigg\}\nn\\
&&+\,h.c.\,,
\qquad\label{eq:HeffKLmue}
\eea
where
\bea\label{eq:R}
R(z_i,z_j,y_k,y_l) &=& F_H(z_i,y_l)+ F_H(z_j,y_k)- F_H(z_j,y_l)- F_H(z_i,y_k)\,.\eea

We remark that the relevant operators differ from  $(\bar d\gamma_\mu e)(\bar \mu\gamma^\mu s)$ present in the PS model which is  characteristic for transitions mediated by leptoquark exchanges. Therefore, the branching ratio for $K_L\to\mu e$ in the LHT model is most straightforwardly obtained from the one for $K_L\to\mu^+\mu^-$ in the SM, after the difference between $\mu^+\mu^-$ and $\mu e$ has been taken into account.

With the help of (XI.44) and (XXV.1) of \cite{BBL} we easily find
\bea\label{eq:BrKLmue}
Br(K_L\to\mu e)& =&Br(K_L\to\mu^+ e^-)+Br(K_L\to\mu^- e^+)
\\
&=&\frac{G_F^2}{128\pi^4}M_{W_L}^4\frac{v^4}{f^4}
Br(K^+\to\mu^+\nu) \frac{\tau(K_L)}{\tau(K^+)}\frac{1}{|V_{us}|^2}\nn\\
&&\hspace{1.5cm}
\cdot \left|\sum_{i,j=2,3}\text{Re}(\xi^{(K)}_i) \chi^{(\mu e)}_j R(z_i,z_1,y_1,y_j) \right|^2\,.
\eea
Here \cite{PDG,CKMreport}
\be
Br(K^+\to\mu^+\nu)=(63.44 \pm 0.14)\%\,,\quad
\frac{\tau(K_L)}{\tau(K^+)}=4.117 \pm 0.019\,,\quad |V_{us}|=0.225 \pm 0.001\,.
\ee

The corresponding expression for $Br(K_S\to\mu e)$ is obtained from \eqref{eq:BrKLmue} through the replacements \cite{PDG}
\be\label{eq:BrKSmue}
\frac{\tau(K_L)}{\tau(K^+)}\longrightarrow \frac{\tau(K_S)}{\tau(K^+)}=(7.229\pm0.014)\cdot10^{-3}\,, \qquad \text{Re}(\xi_i^{(K)}) \longrightarrow \text{Im}(\xi_i^{(K)})\,.
\ee
Due to $\tau(K_S)\ll \tau(K_L)$, the branching ratio $Br(K_S\to\mu e)$ is expected to be typically by two orders of magnitude smaller than $Br(K_L\to\mu e)$ unless $\text{Im}(\xi_i^{(K)})\gg \text{Re}(\xi_i^{(K)})$.

The decay $K_L\to\pi^0\mu e$ in the LHT model is again governed by the effective Hamiltonian in \eqref{eq:HeffKLmue}. This time it is useful to perform the calculation of the branching ratio  in analogy with $\klpn$ \cite{LesHouches}. Removing the overall factor 3 in $Br(\klpn)$ corresponding to three neutrino flavours, we find
\bea
Br(K_L\to\pi^0\mu e)&=&Br(K_L\to\pi^0\mu^+ e^-)+Br(K_L\to\pi^0\mu^- e^+)\nn\\
&=& \frac{G_F^2 M_{W_L}^4}{128\pi^4}\frac{v^4}{f^4}Br(K^+\to\pi^0\mu^+\nu)\frac{\tau(K_L)}{\tau(K^+)}\frac{1}{|V_{us}|^2}\nn\\
&&\hspace{1.5cm}\cdot \left| \sum_{i,j=2,3}\text{Im}(\xi_i^{(K)})\chi_j^{(\mu e)} R(z_i,z_1,y_1,y_j) \right|^2\,.\label{eq:BrKLpi0mue}
\eea
Here \cite{PDG},
\be
Br(K^+\to\pi^0\mu^+\nu)=(3.32\pm0.06)\%\,.
\ee
We note that this time $\text{Im}(\xi_i^{(K)})$ instead of $\text{Re}(\xi_i^{(K)})$ enters, coming from the difference in sign between the relations
\be
\langle \pi^0 | (\bar ds)_{V-A} | \bar K^0\rangle = -
\langle \pi^0 | (\bar sd)_{V-A} |  K^0\rangle
\ee
and
 \be
\langle 0 | (\bar ds)_{V-A} |\bar  K^0\rangle = +
\langle 0 | (\bar sd)_{V-A} |  K^0\rangle\,.
\ee

The corresponding expression for $Br(K_S\to\pi^0\mu e)$ is obtained from \eqref{eq:BrKLpi0mue} through the replacements
\be\label{eq:BrKSpi0mue}
\frac{\tau(K_L)}{\tau(K^+)}\longrightarrow \frac{\tau(K_S)}{\tau(K^+)}\,, \qquad \text{Im}(\xi_i^{(K)}) \longrightarrow \text{Re}(\xi_i^{(K)})\,.
\ee

We would like to emphasize that in deriving the formulae for $Br(K_{L,S}\to\mu e)$ and $Br(K_{L,S}\to\pi^0\mu e)$ we have neglected the contributions from CP violation in $K^0-\bar K^0$ mixing (indirect CP violation). For instance in the case of $Br(K_L\to\pi^0\mu e)$ only CP violation in the amplitude (direct CP violation) has been taken into account. The indirect CP violation alone gives the contribution
\be
Br(K_L\to\pi^0\mu e)^\text{ind.}=Br(K_S\to\pi^0\mu e) |\eps_K|^2
\ee
and needs only to be taken into account, together with the interference with
the contribution from direct CP violation, for $\text{Im}(\xi_i^{(K)}) \ll
\text{Re}(\xi_i^{(K)})$. {The latter case, however, is uninteresting since it
corresponds to an unmeasurably small branching ratio. This should be
contrasted with the case of $K_L\to\pi^0e^+e^-$, where the indirect CP
violation turns out to be dominant \cite{Buchalla:2003sj}.} The origin of the difference is that photon penguins, absent in $K_L\to\pi^0\mu e$, are present in  $K_L\to\pi^0e^+e^-$ and the structure of $\Heff$ is rather different from \eqref{eq:HeffKLmue}. Moreover, while the estimate of indirect CP violation to $K_L\to\pi^0\mu e$ in the LHT model can be done perturbatively, this is not the case for $K_L\to\pi^0e^+e^-$, where the matrix elements of the usual $\Delta F=1$ four quark operators have to be taken into account together with renormalization group effects at scales below $M_W$. 

In summary the branching ratios for $Br(K_{L,S}\to\mu e)$ and  $Br(K_{L,S}\to\pi^0\mu e)$ in the LHT model can be calculated fully in perturbation theory and are thus as theoretically clean as $\klpn$. As seen in the formulae \eqref{eq:BrKLmue}, \eqref{eq:BrKSmue}, \eqref{eq:BrKLpi0mue} and \eqref{eq:BrKSpi0mue} above, $Br(K_L\to\mu e)$ and $Br(K_S\to\pi^0\mu e)$ are governed by $\text{Re}(\xi_i^{(K)})$, while $Br(K_S\to\mu e)$ and $Br(K_L\to\pi^0\mu e)$ by $\text{Im}(\xi_i^{(K)})$. Neglecting the indirect CP violation, we find then that in the so-called ``$K$-scenario'' of \cite{Blanke:2006eb}, the T-odd contributions to $Br(K_L\to\mu e)$ and $Br(K_S\to\pi^0\mu e)$ are highly suppressed, while $Br(K_S\to\mu e)$ and $Br(K_L\to\pi^0\mu e)$ are generally non-vanishing. The opposite is true in the case of the ``$B_s$-scenario'' in which only $Br(K_L\to\mu e)$ and $Br(K_S\to\pi^0\mu e)$ differ significantly from zero.

This discussion shows that the measurements of $Br(K_L\to\mu e)$ and $Br(K_L\to\pi^0\mu e)$ will transparently shed some light on the complex phases present in the mirror quark sector, and from the point of view of the LHT model, the measurement of $Br(K_L\to\pi^0\mu e)$ at the level of $10^{-15}$ will be a clear signal of new CP-violating phases at work.

\boldmath
\newsection{$B_{d,s} \rightarrow \mu e$, $B_{d,s} \rightarrow \tau e$ and $B_{d,s} \rightarrow \tau \mu$}\label{sec:Bsd}
\unboldmath

The decays of neutral $B$-mesons to two different charged leptons proceed
similarly to the $K_L \rightarrow \mu e$ decay discussed in Section
\ref{sec:KLmue}.
The effective Hamiltonian describing these processes receive again contributions
only from box diagrams.
For the $B_d \rightarrow \mu e$ decay, it reads
\bea 
\Heff(B_d \rightarrow \mu e)&=&\frac{G_F^2}{32\pi^2}M_{W_L}^2\frac{v^2}{f^2}\sum_{i,j}\xi_i^{(d)}
F_H(z_i,y_j)\left[\chi^{(\mu e)}_j(\bar b d)_{V-A}(\bar
  e\mu)_{V-A}\right. \nn\\
&&\hspace{1.5cm}\left. + \chi^{(\mu e)*}_j(\bar b d)_{V-A}(\bar \mu e)_{V-A}\right]\,,
\eea
with $\xi_i^{(d)}=V_{Hd}^{ib*} V_{Hd}^{id}$.
Using the
unitarity of the $V_{Hd}$ and $V_{H \ell}$ matrices, it becomes
\bea
\Heff(B_d \rightarrow \mu e) &=&
\frac{G_F^2}{32\pi^2}M_{W_L}^2\frac{v^2}{f^2}
\sum_{i,j=2,3}\xi_i^{(d)}R(z_i,z_1,y_1,y_j)
\left[
  \chi_j^{(\mu e)} (\bar b d)_{V-A}(\bar e\mu)_{V-A}\right. \nn\\
&&\hspace{1.5cm}\left. + 
  \chi_j^{(\mu e)*} (\bar b d)_{V-A}(\bar \mu e)_{V-A} 
\right]\,,\qquad\label{eq:HeffBdmue}
\eea
with $R(z_i,z_j,y_k,y_l)$ being the combination of short distance functions
defined in (\ref{eq:R}).
 
The effective Hamiltonians describing the remaining decays have a similar
structure and can easily be derived from $\Heff(B_d \rightarrow \mu
e)$ in (\ref{eq:HeffBdmue}) through the following replacements
\bea
\Heff(B_s \rightarrow \mu e)&:&\qquad \xi_i^{(d)} \rightarrow
\xi_i^{(s)}\,,\nn\\
\Heff(B_d \rightarrow \tau e)&:&\qquad \chi_j^{(\mu e)} \rightarrow
\chi_j^{(\tau e)}\,,\nn\\
\Heff(B_s \rightarrow \tau e)&:&\qquad \xi_i^{(d)} \rightarrow
\xi_i^{(s)}\,,\chi_j^{(\mu e)} \rightarrow
\chi_j^{(\tau e)}\,,\nn\\
\Heff(B_d \rightarrow \tau \mu)&:&\qquad \chi_j^{(\mu e)} \rightarrow
\chi_j^{(\tau \mu)}\,,\nn\\
\Heff(B_s \rightarrow \tau \mu)&:&\qquad \xi_i^{(d)} \rightarrow
\xi_i^{(s)}\,,\chi_j^{(\mu e)} \rightarrow
\chi_j^{(\tau \mu)}\,.\label{eq:HeffBds}
\eea
The effective Hamiltonians for the $\bar B_{d,s}$ decays are simply given by the hermitian conjugates of the corresponding expressions in \eqref{eq:HeffBdmue} and \eqref{eq:HeffBds}.

In calculating the corresponding branching ratios, it is important to observe
that while in $K_L \rightarrow \mu e$ the decaying meson, $K_L \simeq (\bar s
d + \bar d s)/\sqrt{2}$, is a mixture of flavour eigenstates, the decaying
$B_{d,s}$ mesons are instead flavour eigenstates (e.\,g. $B_d=\bar b d$).
For this reason, in $Br(K_L \rightarrow \mu e)$ the two conjugate
contributions of $\Heff$ combine together, while in $B_{d,s}$ decays
only one contribution enters, with the conjugate one describing $\bar B_{d,s}$
decays.
With this consideration in mind, starting from 
$Br(K_L \rightarrow \mu e)$ in (\ref{eq:BrKLmue}), one can straightforwardly
find the following expressions
\bea
Br(\bar B_d \rightarrow \mu e) &=& Br(B_d \rightarrow \mu e) = Br(B_d
\rightarrow \mu^+ e^-) + Br(B_d \rightarrow \mu^- e^+)  \nn\\
&=&\frac{G_F^2
  M_{W_L}^4}{512\pi^4|V_{ub}|^2}\frac{v^4}{f^4}\frac{\tau(B_d)}{\tau(B^+)} Br(B^+ \rightarrow \mu^+ \nu_\mu)\\
&&\cdot
 \left[\left|\sum_{i,j=2,3}
    \xi_i^{(d)} \chi_j^{(\mu e)} R(z_i,z_1,y_1,y_j)\right|^2
+\left|\sum_{i,j=2,3}
    \xi_i^{(d)} \chi_j^{(\mu e)*} R(z_i,z_1,y_1,y_j)\right|^2\right] \nn \,,\\\nn\\
Br(\bar B_s \rightarrow \mu e) &=& Br(B_s \rightarrow \mu e) = Br(B_s
\rightarrow \mu^+ e^-) + Br(B_s \rightarrow \mu^- e^+)  \nn\\
&=&\frac{G_F^2
  M_{W_L}^4}{512\pi^4|V_{ub}|^2}\frac{v^4}{f^4}\frac{\tau(B_s)}{\tau(B^+)}\frac{M_{B_s}}{M_{B_d}}\frac{F_{B_s}^2}{F_{B_d}^2} Br(B^+ \rightarrow \mu^+ \nu_\mu)\\\
&&\cdot 
 \left[\left|\sum_{i,j=2,3}
    \xi_i^{(s)} \chi_j^{(\mu e)} R(z_i,z_1,y_1,y_j)\right|^2 +\left|\sum_{i,j=2,3}
    \xi_i^{(s)} \chi_j^{(\mu e)*} R(z_i,z_1,y_1,y_j)\right|^2\right] \,,\nn \\\nn\\
Br(\bar B_d \rightarrow \tau e) &=& Br(B_d \rightarrow \tau e) = Br(B_d
\rightarrow \tau^+ e^-) + Br(B_d \rightarrow \tau^- e^+)  \nn\\
&=&\frac{G_F^2
  M_{W_L}^4}{512\pi^4|V_{ub}|^2}\frac{v^4}{f^4}\frac{\tau(B_d)}{\tau(B^+)} Br(B^+ \rightarrow \tau^+ \nu_\tau)\\
&&\cdot 
 \left[\left|\sum_{i,j=2,3}
    \xi_i^{(d)} \chi_j^{(\tau e)} R(z_i,z_1,y_1,y_j)\right|^2
+\left|\sum_{i,j=2,3}
    \xi_i^{(d)} \chi_j^{(\tau e)*} R(z_i,z_1,y_1,y_j)\right|^2\right] \,,\nn\\\nn\\
Br(\bar B_s \rightarrow \tau e) &=& Br(B_s \rightarrow \tau e) = Br(B_s
\rightarrow \tau^+ e^-) + Br(B_s \rightarrow \tau^- e^+)  \nn\\
&=&\frac{G_F^2
  M_{W_L}^4}{512\pi^4|V_{ub}|^2}\frac{v^4}{f^4}\frac{\tau(B_s)}{\tau(B^+)}\frac{M_{B_s}}{M_{B_d}}\frac{F_{B_s}^2}{F_{B_d}^2} Br(B^+ \rightarrow \tau^+ \nu_\tau)\\
&&\cdot
 \left[\left|\sum_{i,j=2,3}
    \xi_i^{(s)} \chi_j^{(\tau e)} R(z_i,z_1,y_1,y_j)\right|^2+\left|\sum_{i,j=2,3}
    \xi_i^{(s)} \chi_j^{(\tau e)*} R(z_i,z_1,y_1,y_j)\right|^2\right] \,,\nn\\\nn\\
Br(\bar B_d \rightarrow \tau \mu) &=& Br(B_d \rightarrow \tau \mu) = Br(B_d
\rightarrow \tau^+ \mu^-) + Br(B_d \rightarrow \tau^- \mu^+)  \nn\\
&=&\frac{G_F^2
  M_{W_L}^4}{512\pi^4|V_{ub}|^2}\frac{v^4}{f^4}\frac{\tau(B_d)}{\tau(B^+)} Br(B^+ \rightarrow \tau^+ \nu_\tau)\\
&&\cdot 
 \left[\left|\sum_{i,j=2,3}
    \xi_i^{(d)} \chi_j^{(\tau \mu)} R(z_i,z_1,y_1,y_j)\right|^2
+\left|\sum_{i,j=2,3}
    \xi_i^{(d)} \chi_j^{(\tau \mu)*} R(z_i,z_1,y_1,y_j)\right|^2\right] \,,\nn\\\nn\\
Br(\bar B_s \rightarrow \tau \mu) &=& Br(B_s \rightarrow \tau \mu) = Br(B_s
\rightarrow \tau^+ \mu^-) + Br(B_s \rightarrow \tau^- \mu^+)  \nn\\
&=&\frac{G_F^2
  M_{W_L}^4}{512\pi^4|V_{ub}|^2}\frac{v^4}{f^4}\frac{\tau(B_s)}{\tau(B^+)}\frac{M_{B_s}}{M_{B_d}}\frac{F_{B_s}^2}{F_{B_d}^2} Br(B^+ \rightarrow \tau^+ \nu_\tau)\\
&&\cdot
 \left[\left|\sum_{i,j=2,3}
    \xi_i^{(s)} \chi_j^{(\tau \mu)} R(z_i,z_1,y_1,y_j)\right|^2
+\left|\sum_{i,j=2,3}
    \xi_i^{(s)} \chi_j^{(\tau \mu)*} R(z_i,z_1,y_1,y_j)\right|^2\right] \,.\nn
\eea 
Analogously to $Br(K_L \rightarrow \mu e)$, we have chosen to normalize the
branching ratios in question introducing the branching ratio of a $B^+$
leptonic decay.
This normalization will become helpful once $Br(B^+ \rightarrow \mu^+
\nu_\mu)$ and $Br(B^+ \rightarrow \tau^+ \nu_\tau)$ are experimentally measured with sufficient accuracy. Currently the measurements for   $Br(B^+ \rightarrow \tau^+ \nu_\tau)$ read
\be
Br(B^+\to\tau^+\nu_\tau)=\left\{\begin{array}{l}\left(1.79^{+0.56+0.39}_{-0.49-0.46}\right)\cdot 10^{-4} \qquad\text{\cite{btau-BELLE}}\,, \\
\left(0.88^{+0.68+0.11}_{-0.67-0.11}\right)\cdot10^{-4} \qquad\text{\cite{btau-BABAR}}\,,
\end{array}\right.
\ee
while the experimental upper bound on $Br(B^+ \rightarrow \mu^+
\nu_\mu)$ is given by \cite{bmu}
\be
Br(B^+ \rightarrow \mu^+
\nu_\mu)<6.6\cdot10^{-6}\qquad(90\%\,\text{C.L.})\,.
\ee
Therefore, in our numerical analysis in Section \ref{sec:num} we will use the central values of the theoretical predictions for these decays, given by
\bea
Br(B^+ \rightarrow \mu^+ \nu_\mu) &=& \frac{G_F^2}{8\pi}|V_{ub}|^2F_{B_d}^2M_{B_d}m_\mu^2\tau{(B^+)}=(3.8\pm1.1)\cdot10^{-7}\,,\\
Br(B^+\to\tau^+\nu_\tau) &=& \frac{G_F^2}{8\pi}|V_{ub}|^2F_{B_d}^2M_{B_d}m_\tau^2\tau{(B^+)}={(1.1\pm0.3)\cdot10^{-4}}\,,
\eea
where the relevant input parameters are collected in Table \ref{tab:input} of Section \ref{sec:num}.

\boldmath
\newsection{$\tau^-\to e^-\mu^+e^-$ and $\tau^-\to\mu^- e^+\mu^-$}
\unboldmath\label{sec:teme}

\begin{figure}
\center{\epsfig{file=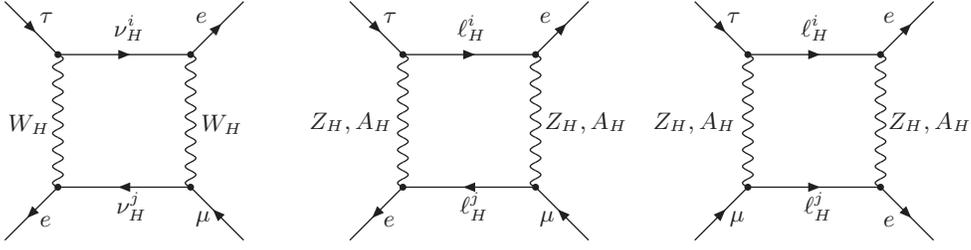}}
\caption{\it Diagrams contributing to $\tau^-\to e^-\mu^+e^-$ in the LHT model.\label{fig:teme}}
\end{figure}

These two decays are of $\Delta L=2$ type and are very strongly suppressed in the SM. In the LHT model they proceed through the box diagrams in Fig.~\ref{fig:teme}.

The effective Hamiltonians for these decays can be obtained from $\Delta B=2$ processes, that is from (3.11) of \cite{BBPTUW}. In the case of $\tau^-\to e^-\mu^+e^-$ we find
\be\label{eq:Heffteme}
\Heff = \frac{G_F^2}{16\pi^2}M_{W_L}^2\frac{v^2}{f^2} \sum_{i,j}\chi_i^{(\tau e)}\chi_j^{(\mu e)} F_H(y_i,y_j)(\bar e \tau)_{V-A}(\bar e\mu)_{V-A}
\ee
with the function $F_H$ given in \eqref{eq:FH}. The additional factor 4 relative to (3.11) of \cite{BBPTUW} results from the different flavour structure of the operator involved and the two identical particles in the final state.

The corresponding effective Hamiltonian for $\tau^-\to\mu^- e^+\mu^-$ is obtained by simply exchanging $e$ and $\mu$, with $\chi_j^{(e\mu)}={\chi_j^{(\mu e)}}^*$.

The relevant branching ratios can be found by comparing these two decays to the tree level decay $\tau^-\to \nu_\tau e^-\bar\nu_e$, for which the effective Hamiltonian reads
\be
\Heff= \frac{G_F}{\sqrt{2}} (\bar\nu_\tau\tau)_{V-A}(\bar e\nu_e)_{V-A}\,,
\ee
{and yields the decay rate}
\be
\Gamma(\tau^-\to \nu_\tau e^-\bar\nu_e)=\frac{G_F^2m_\tau^5}{192\pi^3}\,.
\ee

For $\tau^-\to e^-\mu^+e^-$ we find then 
\be\label{eq:Brteme}
Br(\tau^-\to e^-\mu^+e^-)=\frac{m_\tau^5\tau_\tau}{192\pi^3}\left(\frac{G_F^2 M_{W_L}^2}{16\pi^2}\right)^2\frac{v^4}{f^4}
\left|\sum_{i,j} \chi_i^{(\tau e)}\chi_j^{(\mu e)} F_H(y_i,y_j)\right|^2,
\ee
{where  we neglected $m_{\mu,e}$ with respect to $m_\tau$.} We have included the factor $1/2$ to take into account the presence of two identical fermions in the final state.

The branching ratio for $\tau^-\to\mu^-e^+\mu^-$ is obtained from \eqref{eq:Brteme} by interchanging $\mu\leftrightarrow e$.

\boldmath
\newsection{$\tau^-\to\mu^-e^+e^-$ and $\tau^-\to e^-\mu^+\mu^-$}
\unboldmath\label{sec:tmee}

These decays have two types of contributions. First of all they proceed as in $\tau^-\to\mu^-\mu^+\mu^-$ and  $\tau^-\to e^-e^+e^-$ through $\Delta L=1$ penguin and box diagrams. As this time there are no identical particles in the final state, the effective Hamiltonians for these contributions can be directly obtained from the  decay $B\to X_s\ell^+\ell^-$.
Let us derive explicitly the effective Hamiltonian for $\tau^-\to\mu^-e^+e^-$. The generalization to $\tau^-\to e^-\mu^+\mu^-$ will then be automatic.

As the QCD corrections are not involved now, only three operators originating in magnetic photon penguins, $Z^0$-penguins, standard photon penguins and the relevant box diagrams have to be considered.  Keeping the notation from $B\to X_s\mu^+\mu^-$ but translating the quark flavours into lepton flavours these operators are
\be
\mathcal{Q}_7=\frac{e}{8\pi^2}m_\tau \bar \mu \sigma^{\alpha\beta}(1+\gamma_5)\tau F_{\alpha\beta}\,,
\ee
 {that enters}, of course with different external states, also the $\mu\to e\gamma$ decay, and
\be
\mathcal{Q}_9=(\bar \mu\tau)_{V-A}(\bar ee)_V\,,\qquad
\mathcal{Q}_{10}=(\bar \mu\tau)_{V-A}(\bar ee)_A\,.
\ee
The effective Hamiltonian is then given by
\be
\Heff(\tau^-\to\mu^-e^+e^-)=-\frac{G_F}{\sqrt{2}}\left[C_7^{\tau\mu} \mathcal{Q}_7+C_9^{\tau\mu}  \mathcal{Q}_9 +C_{10}^{\tau\mu} \mathcal{Q}_{10}\right]\,.
\ee

The Wilson coefficient for the operator $\mathcal{Q}_7$ can easily be found from Section~\ref{sec:liljgamma} of the present paper and Section 7 of \cite{Blanke:2006eb}. We find
\be\label{eq:C7}
C_7^{\tau\mu}=-\frac{1}{2}\bar D^{\prime\,\tau\mu}_\text{odd}\,,
\ee
with $\bar D^{\prime\,\tau\mu}_\text{odd}$ obtained from \eqref{eq:Dpoddme} by replacing $(\mu e)$ with $(\tau\mu)$.

The Wilson coefficients of the operators $\mathcal{Q}_9$ and $\mathcal{Q}_{10}$ receive not only contributions from $\Delta L=1$ $\gamma$-penguin, $Z^0$-penguin and box diagrams, but also from $\Delta L=2$ box diagrams, as in Section \ref{sec:teme}. For $C_9^{\tau\mu}$ and $C_{10}^{\tau\mu}$ we can then write
\begin{gather}
C_9^{\tau\mu}=\frac{\alpha}{2\pi}\tilde C_9^{\tau\mu}\,,\qquad C_{10}^{\tau\mu}=\frac{\alpha}{2\pi}\tilde C_{10}^{\tau\mu}\,,\\
\label{eq:C9}
\tilde C_9^{\tau\mu}=\frac{\bar Y_{e,\text{odd}}^{\tau\mu}}{\sin^2\theta_W}-4\bar Z_\text{odd}^{\mu e}-\Delta_{\tau\mu}\,,\qquad
\tilde C_{10}^{\tau\mu}=-\frac{\bar Y_{e,\text{odd}}^{\tau\mu}}{\sin^2\theta_W}+\Delta_{\tau\mu}\,,
\end{gather}
with the functions $\bar Y_{e,\text{odd}}^{\tau\mu}$ and $\bar Z_{\text{odd}}^{\tau\mu}$ obtained from \eqref{eq:Yoddme} and  \eqref{eq:Zoddme}   by replacing ($\mu e$) by ($\tau\mu$). 
$\Delta_{\tau\mu}$ represents the additional $\Delta L=2$ contribution which is not present in the case of $b\to s\ell^+\ell^-$ and will be explained below.
As there are no light fermions in the T-odd sector, the mass independent term present in $C_9$ in the case of $b\to s\ell^+\ell^-$ in (X.5) of \cite{BBL} is absent here. Effectively this corresponds to setting $\eta=1$ in the latter equation and of course removing QCD corrections.

\begin{figure}
\center{\epsfig{file=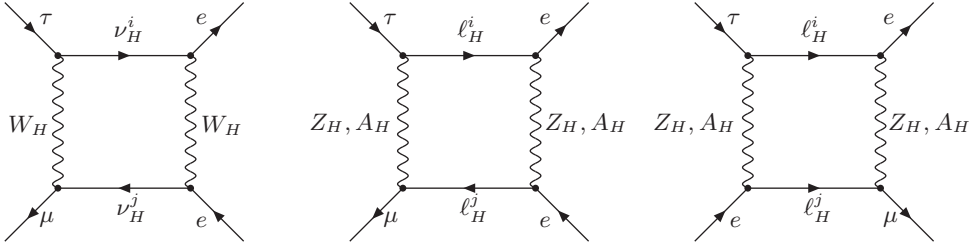}}
\caption{\it Diagrams of $\Delta L=2$ type contributing to $\tau^-\to \mu^-e^+e^-$ in the LHT model.\label{fig:tmee}}
\end{figure}

As already mentioned, also the $\Delta L=2$ diagrams shown in  Fig.~\ref{fig:tmee} contribute to this decay. The corresponding effective Hamiltonian can be obtained from \eqref{eq:Heffteme} through the obvious replacements of local operators, removing the symmetry factor 2 and the following change in the mixing factors:
\be
 \chi_i^{(\tau e)} {\chi_j^{(\mu e)}}\longrightarrow  \chi_i^{(\tau e)} {\chi_j^{(\mu e)}}^*\,, 
\ee
so that we find
\bea
\Delta_{\tau\mu} &=& \frac{2\pi}{\alpha}\frac{G_F}{32\pi^2}\sqrt{2}M_{W_L}^2\frac{v^2}{f^2}\sum_{i,j}\chi_i^{(\tau e)}{\chi_j^{(\mu e)}}^*F_H(y_i,y_j)\nn\\
&=& \frac{1}{16\sin^2\theta_W}\frac{v^2}{f^2}\sum_{i,j}\chi_i^{(\tau e)}{\chi_j^{(\mu e)}}^*F_H(y_i,y_j) \,.
\eea

Effectively the presence of the diagrams in Fig.~\ref{fig:tmee} introduces corrections to the Wilson coefficients $\tilde C_9$ and $\tilde C_{10}$ in \eqref{eq:C9}. As the relevant operator has the structure $(V-A)\otimes(V-A)$, the shifts in $\tilde C_9$ and $\tilde C_{10}$ are equal up to an overall sign.

Finally, introducing

\be\label{eq:Rsdef}
\hat s=\frac{(p_{e^+}+p_{e^-})^2}{m_\tau^2}\,,\qquad 
R^{\tau\mu}(\hat s)=\frac{\frac{d}{d\hat s}{\Gamma(\tau^-\to\mu^- e^+e^-)}}{\Gamma(\tau^-\to \mu^-\bar\nu_\mu\nu_\tau)}
\ee

and neglecting $m_e$ with respect to $m_\tau$ we find for the differential decay rate $R^{\tau\mu}(\hat s)$
\bea\label{eq:Rs}
R^{\tau\mu}(\hat s)&=&\frac{\alpha^2}{4\pi^2}(1-\hat s)^2
\bigg[(1+2\hat s)\left(|\tilde{C}_9^{\tau\mu}|^2+|\tilde C_{10}^{\tau\mu}|^2\right)\nn\\
&& \hspace{2.5cm}
+4\left(1+\frac{2}{\hat s}\right)|C_7^{\tau\mu}|^2+12\,\text{Re}\left(C_7^{\tau\mu}{(\tilde{C}_9^{\tau\mu})}^*\right)
\bigg]\,.
\eea
The branching ratio is then given as follows:
\be\label{eq:Brtmee}
Br(\tau^-\to\mu^-e^+e^-)=Br(\tau^-\to \mu^-\bar\nu_\mu\nu_\tau)\int_{4m_e^2/m_\tau^2}^1 R^{\tau\mu}(\hat s)\,d\hat s\,.
\ee

The branching ratio for $\tau^-\to e^-\mu^+\mu^-$ can easily be obtained from the above expressions by interchanging $\mu\leftrightarrow e$, where $\chi_i^{(e\mu)}={\chi_i^{(\mu e)}}^*$.

 For quasi-degenerate mirror leptons the $\Delta L=1$ part clearly dominates as the GIM-like suppression acts only on one mirror lepton propagator, whereas it acts twice in the $\Delta L=2$ case. Moreover, in the latter case the effective Hamiltonian is quartic in the $V_{H\ell}$ couplings, whereas it is to a very good approximation quadratic in the case of $\Delta L=1$. As these factors are all smaller than 1, quite generally $\Delta L=2$ contributions will then be additionally suppressed by the mixing matrix elements. Consequently, the $\Delta L=1$ part is expected to dominate and the shift $\Delta_{\tau\mu}$ can be neglected. 
On the other hand, for very special structures of the $V_{H\ell}$ matrix, the double GIM suppression of $\Delta L=2$ with respect to $\Delta L=1$ contributions could be compensated by the $V_{H\ell}$ factors. Therefore it is safer to use the more general expressions given above.

\newsection{$\bm{(g-2)_\mu}$ in the LHT Model}\label{sec:g-2}

The anomalous magnetic moment of the muon $a_{\mu}=(g-2)_\mu/2$ provides an
excellent test for physics beyond the SM and has been measured very precisely
at the  E821 experiment \cite{Bennett:2006fi} in Brookhaven. The latest result
of the ${(g-2)}$ Collaboration of E821 reads
\begin{equation}
a_{\mu}^{\rm{exp}} = (11659208.0 \pm 6.3) \cdot 10^{-10}\,, \label{Exp}
\end{equation}
whereas the SM prediction is given by \cite{Hagiwara:2003da}
\begin{equation}
a_{\mu}^{\rm{SM}} ={a_{\mu}^{\rm{QED}} + a_{\mu}^{\rm{ew}} + a_{\mu}^{\rm{had}}}
= (11659180.4 \pm 5.1) \cdot 10^{-10}\,. \label{SM}
\end{equation}
While the QED and electroweak contributions to $a_\mu^\text{SM}$ are known very precisely \cite{QEDg-2,EWg-2}, the theoretical uncertainty is dominated by the hadronic vacuum polarization and light-by-light contributions. These contributions have been evaluated in \cite{Hagiwara:2003da,Davier:2003pw,deTroconiz:2004tr,LBL}.

The anomalous magnetic moment $a_{\mu}$ can be extracted from the photon-muon vertex function $\Gamma^{\mu}(p^{\prime},p)$
\begin{equation}
\bar{u}(p^{\prime}) \Gamma^{\mu}(p^{\prime},p) u(p)=
\bar{u}(p^{\prime})\left[\gamma^{\mu} F_{V}(q^{2}) + (p+p^{\prime})^{\mu} F_{M}(q^2)\right]u(p)\,,
\end{equation}
where the anomalous magnetic moment of the muon $a_{\mu}$ can be read off as
\begin{equation}
a_{\mu}=-2mF_{M}(0)\,.
\end{equation}

\begin{figure}
\center{\epsfig{file=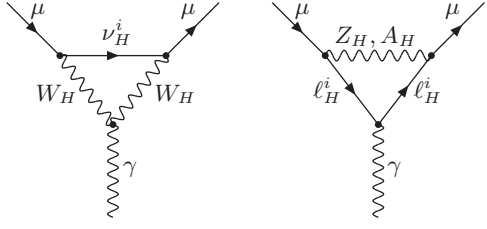}}
\caption{\it Diagrams contributing to $(g-2)_\mu$ in the LHT model.\label{fig:g-2}}
\end{figure}

The diagrams  which yield new contributions to $a_{\mu}$ in the LHT model are shown in Fig.~\ref{fig:g-2}. They either have a heavy neutral gauge boson ($Z_H$ or $A_H$) and two heavy charged leptons $\ell_{H}^{i} \,\, (i=1,2,3)$ or two heavy charged gauge bosons ($W_H^\pm$) and one heavy neutrino $\nu_{H}^{i} \,\, (i=1,2,3)$ running in the loop.

Calculating the diagrams in Fig.~\ref{fig:g-2} and using the {Feynman rules} given in \cite{Blanke:2006eb}, the contributions of the new particles for each generation $i=1,2,3$ are found to be:
\begin{eqnarray}
\left[a_{\mu}\right]^i_{X=A_H,Z_H}&=&\frac{1}{2 \pi^{2}} \frac{m^2_{\mu}}{M^2_{X}} \left|C^i_X\right|^2 r_i \left\{\left(\frac{5}{6}-\frac{5}{2}r_i+r_i^{2}+\left(r_i^{3}-3r_i^{2}+2r_i\right)\ln\frac{r_i-1}{r_i}\right)\right.\nonumber\\
&&\hspace{1cm}+\,\frac{ {m_{Hi}^{\ell}}^2}{2 M_{X}^{2}}\left(\frac{5}{6}+\frac{3}{2}r_i+r_i^{2}+\left(r_i^{2}+r_i^{3}\right)\ln\frac{r_i-1}{r_i}\right)\bigg\}\,,\label{twofermions}\\
\left[a_{\mu}\right]^i_{X=W_H}&=&-\frac{1}{4 \pi^{2}} \frac{m^2_{\mu}}{M^2_{X}}  \left|C^i_X\right|^2 r_i \left\{
-2\left(\frac{5}{6}-\frac{3}{2}b_i+b_i^{2}+\left(b_i^{2}-b_i^{3}\right)\ln\frac{b_i+1}{b_i}\right)\right.\nonumber\\
&&\hspace{1cm}-\,\frac{{m_{Hi}^{\ell}}^2}{M_{X}^{2}}\left(\frac{5}{6}+\frac{5}{2}b_i+b_i^{2}-\left(2b_i+3b_i^{2}+b_i^{3}\right)\ln\frac{b_i+1}{b_i}\right)\Bigg\}\,,\qquad\label{twobosons}
\end{eqnarray}
where
\begin{equation}
r_i=\left(1-\frac{{m_{Hi}^{\ell}}^{2}}{M_{X}^{2}}\right)^{-1}, \qquad b=\frac{{m_{Hi}^{\ell}}^{2}}{M_{X}^{2}}\,r_i
\end{equation}
and
\begin{equation}
C^i_{A_H}=\frac{g^\prime}{20}\,V_{H\ell}^{i\mu}\,, \qquad C^i_{Z_H}=\frac{g}{4}\,V_{H\ell}^{i\mu}\,, \qquad C^i_{W_H}=\frac{g}{2\sqrt{2}}\, V_{H\ell}^{i\mu}\,.
\end{equation}
The parameter $m_{Hi}^{\ell}$ in (\ref{twofermions}) and (\ref{twobosons}) denotes the mass of the mirror leptons while $M_X$ is the mass of the heavy gauge bosons. 
We expanded our results in the small parameter $m_\mu/M_X$. Our results in (\ref{twofermions}) and (\ref{twobosons}) for the muon anomalous magnetic moment are confirmed by the formulae in \cite{Leveille:1977rc} for general couplings.

Replacing the parameters $r$ and $b$ by the more convenient parameter $y_i$,
defined in \eqref{eq:yi}, {leads us to the following expressions} 
\begin{eqnarray}
\left[a_\mu\right]^{}_{Z_H}&=&\frac{\sqrt{2}G_F}{32\pi^2}\frac{v^2}{f^2}m_\mu^2\sum_{i=1}^{3}{ \left|V_{H\ell}^{i\mu}\right|^2}L_1(y_i)\,,
\label{a_mu(x)Z}\\
\left[a_\mu\right]^{}_{A_H}&=&\frac{\sqrt{2}G_F}{160\pi^2}\frac{v^2}{f^2}m_\mu^2\sum_{i=1}^{3}{ \left|V_{H\ell}^{i\mu}\right|^2}L_1(y'_i)\,,
\label{a_mu(x)A}\\
\left[a_\mu\right]^{}_{W_H}&=&\frac{-\sqrt{2}G_F}{32\pi^2}\frac{v^2}{f^2}m_\mu^2\sum_{i=1}^{3}{\left|V_{H\ell}^{i\mu}\right|^2}L_2(y_i)\,,
\label{a_mu(x)W}
\end{eqnarray}
where the functions $L_1$ and $L_2$ are given in Appendix \ref{sec:app2}.

Our final result  for ${a_\mu}$ in the LHT model therefore is
\begin{equation}
a_\mu=\left[a_\mu\right]_\text{SM}+\frac{\sqrt{2}G_F}{32\pi^2}\frac{v^2}{f^2}m_\mu^2\sum_{i=1}^{3}{\left|V_{H\ell}^{i\mu}\right|^2}\left[L_1(y_i)-L_2(y_i)+\frac{1}{5}L_1(y_i^\prime)\right]\,.
\end{equation}

While we disagree with \cite{Goyal}, we confirm the result of \cite{IndianLFV} except that according to us {the factors $(V_{H\nu})_{2i}^* (V_{H\nu})_{2i}$ and $(V_{H\ell})_{2i}^* (V_{H\ell})_{2i}$ in equations (3.22)--(3.24) of that paper should be replaced by $|V_{H\ell}^{i\mu}|^2$.}

\newsection{Numerical Analysis}\label{sec:num}

\subsection{Preliminaries}

In contrast to rare meson decays, the number of flavour violating decays in
the lepton sector, for which useful constraints exist, is rather
limited. Basically only the constraints on $Br(\mu\to e\gamma)$, $Br(\mu^-\to
e^-e^+e^-)$, $R(\mu\text{Ti}\to e\text{Ti})$ and $Br(K_L\to\mu e)$ can be
mentioned here. The situation may change significantly in the coming years and
the next decade through {the experiments briefly discussed} in the introduction.

\begin{table}
\begin{center}
\begin{tabular}{|l|l|}\hline
$m_e=0.5110\mev$ & $\tau(B_d)/\tau(B^+)=0.934(7)$\\
$m_\mu=105.66\mev$ & $\tau(B_s)=1.466(59)\,\text{ps}$ \\ 
$m_\tau=1.7770(3)\gev$ & $\tau(B^+)=1.638(11)\,\text{ps}$ \\
$\tau_\tau = 290.6(10)\cdot 10^{-3}\,\text{ps}$ & $M_{B_d}=5.2794(5)\gev$\\
$M_W=80.425(38)\gev$ & $M_{B_s}=5.3675(18)\gev$ \quad\cite{PDG}\\\cline{2-2}
$\alpha=1/137$ & $|V_{ub}|=3.68(14)\cdot10^{-3}$ \hfill\cite{UTfits} \\\cline{2-2}
$G_F=1.16637(1)\cdot 10^{-5} \gev^{-2}$\quad & $F_8/F_\pi=1.28$ \hfill(ChPT)\\
$\sin^2\theta_W = 0.23122(15)$ \qquad\cite{PDG} & $F_0/F_\pi=1.18(4)$ \\ \cline{1-1}
$F_{B_d}=189(27)\mev$ & $\theta_8=-22.2(18)^\circ$ \\
$F_{B_s}=230(30)\mev$\hfill \cite{Hashimoto} & $\theta_0=-8.7(21)^\circ$\hfill\cite{Escribano}\\\hline
\end{tabular}
\end{center}
\caption{\it Values of the experimental and theoretical quantities used as input parameters.\label{tab:input}}
\end{table}

In this section we want to analyze numerically various branching ratios that we have calculated in Sections \ref{sec:liljgamma}--\ref{sec:g-2}. In Section \ref{sec:corr} we will extend our numerical analysis by studying various ratios of branching ratios and comparing them with those found in the MSSM.
Our purpose is not to present a very detailed numerical analysis of all
decays, but rather to concentrate on the most interesting ones from the
present perspective and indicate rough upper bounds on all calculated
branching ratios within the LHT model. To this end we will first set $f=1\tev$
and consider three benchmark scenarios for the remaining LHT parameters in
\eqref{2.16}, {as discussed below}.

In Table \ref{tab:input} we collect the values of the input parameters that
enter our numerical analysis. {In order to simplify the analysis, we will set
  all input parameters to their central values. As all parameters, except for the
  decay constants $F_{B_{d,s}}$ {and the $\eta - \eta'$ mixing angles},
  are known with quite high precision, including the error ranges in the
  analysis would amount only {to percent effects} in the observables considered, which is clearly beyond the scope of our analysis.}

\subsection{Benchmark Scenarios}\label{sec:scen}

We will consider the following three scenarios:
\vspace{.3cm}

\textbf{Scenario A (red):}

\noindent In this scenario we will choose
\be
V_{H\ell}=V_\text{PMNS}^\dagger\,,
\ee
so that $V_{H\nu}\equiv \mathbbm{1}$, and mirror leptons have no impact on flavour violating observables in the neutrino sector, such as neutrino oscillations. In particular we set {the PMNS parameters to \cite{PMNSparameters} 
\be
\sin\theta_{12}=\sqrt{0.300}\,,\qquad \sin\theta_{13}=\sqrt{0.030}\,,\qquad \sin\theta_{23}=\frac{1}{\sqrt{2}}\,,\qquad \delta_{13}=65^\circ\,,
\ee
which is consistent with the experimental constraints on the PMNS matrix
\cite{PDG}. As no constraints on the PMNS phases exist, we have taken $\delta_{13}$ to be equal to the CKM phase and set the two Majorana phases to zero.}

Furthermore, we take the mirror lepton masses to lie in the range
{
\be\label{eq:massrange}
300\gev\le m^\ell_{Hi}\le1.5\tev\,, \qquad {(i=1,2,3)}\,.
\ee
}

\textbf{Scenario B (green):}

\noindent Here, we take
\be
V_{H\ell}=V_\text{CKM}\,,
\ee
so that \cite{UTfits}
\begin{gather}
\theta^\ell_{12}=13^\circ\,,\qquad \theta^\ell_{13}=0.25^\circ\,,\qquad \theta^\ell_{23}=2.4^\circ\,,\\
\delta^\ell_{12}=0\,,\qquad \delta^\ell_{13}=65^\circ\,,\qquad \delta^\ell_{13}=0\,,
\end{gather}
and the mirror lepton masses in the range \eqref{eq:massrange}.\vspace{.2cm}

\textbf{Scenario C (blue):}

\noindent
Here we perform a general scan over the whole parameter space, with the only
restriction being the range \eqref{eq:massrange} {for mirror lepton masses}.

At a certain stage we will investigate the dependence on mass splittings in the mirror lepton spectrum.

In the case of $K_L\to\mu e$, $B_{d,s}\to\mu e$ and similar decays of Sections \ref{sec:KLmue} and \ref{sec:Bsd}, {the parameters of the mirror quark sector enter and the constraints from $K$ and $B$ physics, analyzed in \cite{BBPTUW,Blanke:2006eb}, have to be taken into account.}

\boldmath
\subsection{$\mu\to e\gamma$, $\mu^-\to e^-e^+e^-$ and $\mu-e$ Conversion}
\unboldmath

In Fig.~\ref{fig:megm3e} we show the correlation between $\mu\to e\gamma$ and $\mu^-\to e^-e^+e^-$ in the scenarios in question together with the experimental bounds on these decays. We observe:

\begin{figure}
{\begin{minipage}{7.5cm}
\epsfig{file=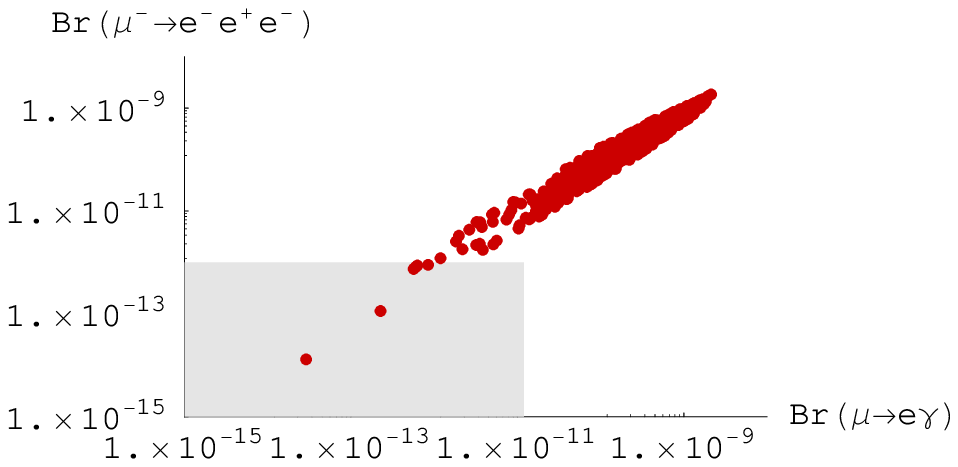,scale=.75}\vspace{-.7cm}
\center{Scenario A}
\end{minipage}}
\begin{minipage}{7.5cm}
\epsfig{file=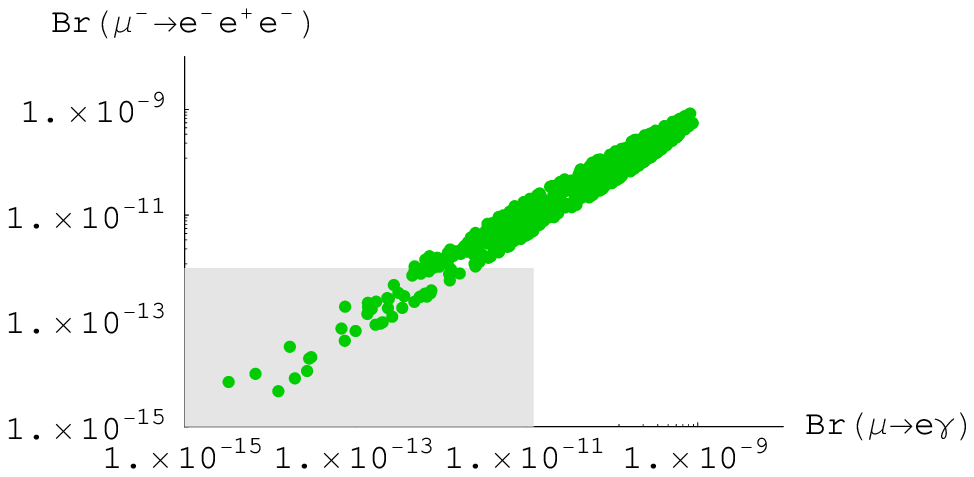,scale=.75}\vspace{-.7cm}
\center{Scenario B}
\end{minipage}\vspace{.8cm}
\begin{minipage}{8.2cm}
\epsfig{file=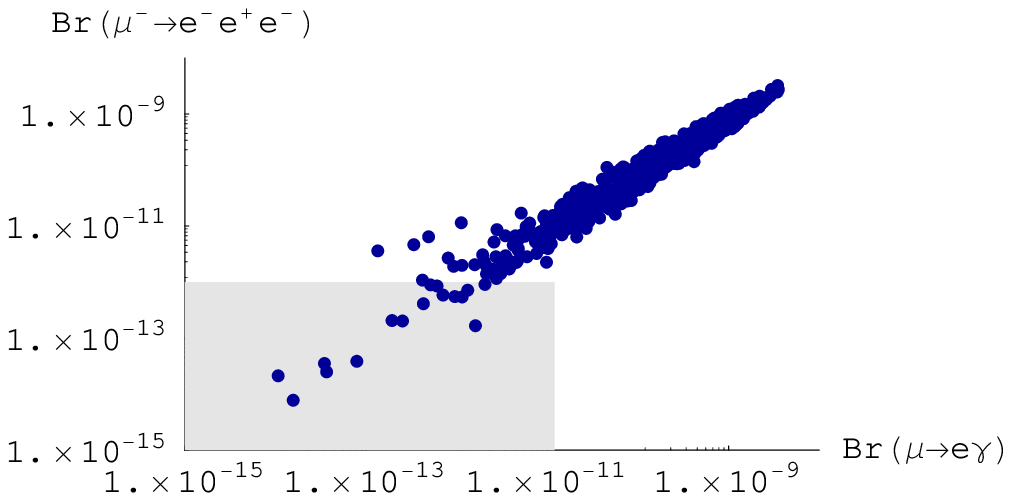,scale=.75}
\end{minipage}
\begin{minipage}{7.5cm}
\epsfig{file=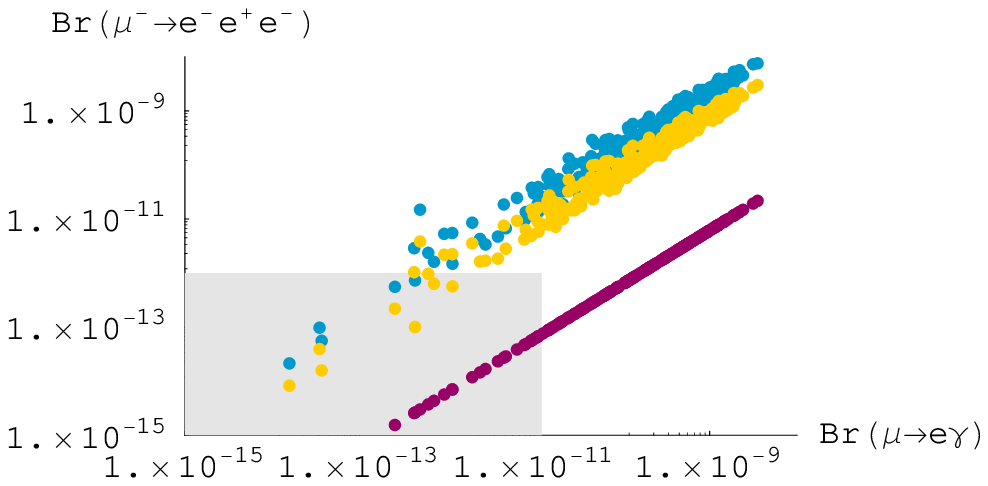,scale=.75}
\end{minipage}\vspace{-.5cm}
\center{Scenario C}\vspace{.5cm}

\caption{\it Correlation between $\mu\to e\gamma$ and $\mu^-\to e^-e^+e^-$ in
  the scenarios of Section~\ref{sec:scen}. In the right plot {of} Scenario C we show the contributions to $\mu^-\to e^-e^+e^-$ from $\bar D_\text{odd}^{\prime\,\mu e}$ (purple, {lowermost}), $\bar Z_\text{odd}^{\mu e}$ (orange, middle) and $\bar Y_{e,\text{odd}}^{\mu e}$ (light-blue, {uppermost}) separately. The shaded area represents the experimental constraints.\label{fig:megm3e}}
\end{figure}

\bi
\item 
In Scenario A the {great majority} of points is outside the allowed range, implying that the $V_{H\ell}$ matrix must be much more hierarchical than $V_\text{PMNS}$ in order to satisfy the present upper bounds on $\mu\to e\gamma$ and $\mu^-\to e^-e^+e^-$.
\item
Also in Scenario B most of the points violate the current experimental bounds, although $V_\text{CKM}$ is much more hierarchical than $V_\text{PMNS}$. The reason is that the CKM hierarchy $s_{13}\ll s_{23} \ll s_{12}$ implies very small effects in transitions between the third and the 
{first generation, like $\tau\to e\gamma$,}
while allowing relatively large effects in the $\mu\to e$ transitions. Thus in order to satisfy the experimental constraints on $\mu\to e\gamma$ and $\mu^-\to e^-e^+e^-$ a very different hierarchy of the $V_{H\ell}$ matrix is required, unless the mirror lepton masses are quasi-degenerate.
\item
In Scenario C there are more possibilities, but also here a strong correlation between $\mu\to e\gamma$ and $\mu^-\to e^-e^+e^-$ is observed. This is easy to understand, as both decays probe dominantly the combinations of $V_{H\ell}$ elements $\chi_i^{(\mu e)}$.
\item 
For Scenario C, we also show the contributions to $\mu^-\to e^-e^+e^-$ from
$\bar D_\text{odd}^{\prime\,\mu e}$, $\bar Z_\text{odd}^{\mu e}$ and $\bar
Y_{e,\text{odd}}^{\mu e}$ separately. We observe that the dominant
contributions come from the functions $\bar Z_\text{odd}^{\mu e}$ and {above all} $\bar Y_{e,\text{odd}}^{\mu e}$, while the contribution of the operator $\mathcal{Q}_7$, given by $\bar D_\text{odd}^{\prime\,\mu e}$, is by roughly two orders of magnitude smaller and thus fully negligible. This should be contrasted with the case of the MSSM where the dipole operator is dominant. We will return to the consequences of this finding in the next section.
\item
We emphasize that the strong correlation between $\mu\to e\gamma$ and
$\mu^-\to e^-e^+e^-$ in the LHT model is not a common feature of all
extensions of the SM, in which the structure of $\mu^-\to e^-e^+e^-$ is
generally much more complicated than in the LHT model. It is clear from
Fig.~\ref{fig:megm3e} that an improved upper bound on $\mu\to e\gamma$ by MEG
in 2007 and in particular its discovery will {provide important information}  on  $\mu^-\to e^-e^+e^-$ within the model in question. 
\ei

\begin{figure}
\center{\epsfig{file=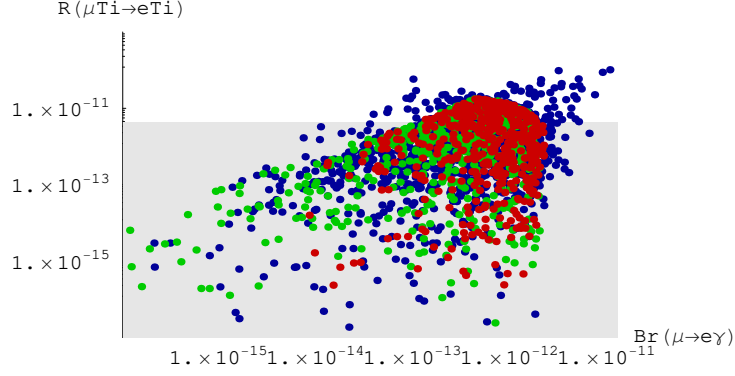,scale=.8}}\vspace{-.4cm}
\caption{\it $\mu-e$ conversion rate in $^{48}_{22}\text{Ti}$ as a function of $Br(\mu\to e\gamma)$, after imposing the existing constraints on $\mu\to e \gamma$ and $\mu^-\to e^-e^+e^-$. The shaded area represents the current experimental upper bound on $R(\mu\text{Ti}\to e\text{Ti})$.\label{fig:mue-conv}}
\end{figure}

Next, in Fig.~\ref{fig:mue-conv} we show the correlation between the $\mu-e$ conversion rate in $^{48}_{22}\text{Ti}$ and $Br(\mu\to e\gamma)$, after imposing the existing constraints on $\mu\to e \gamma$ and $\mu^-\to e^-e^+e^-$. We observe that this correlation is much weaker than the one between $\mu\to e \gamma$ and $\mu^-\to e^-e^+e^-$. Furthermore, we find that the $\mu-e$ conversion rate in Ti is likely to be found close to the current experimental upper bound, and that in some regions of the parameter space the latter bound is even the most constraining one.

\begin{figure}
\begin{minipage}{7.7cm}
\center{\epsfig{file=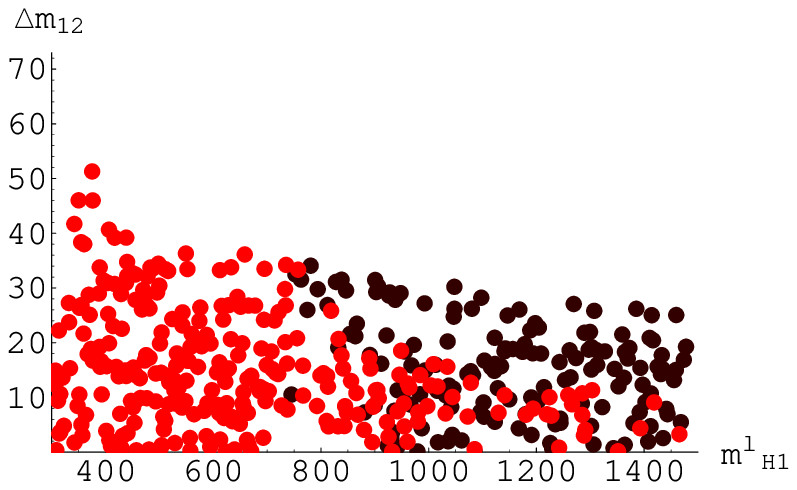,scale=.92}}\vspace{-.2cm}
\center{Scenario A}
\end{minipage}
\begin{minipage}{7.7cm}
\center{\epsfig{file=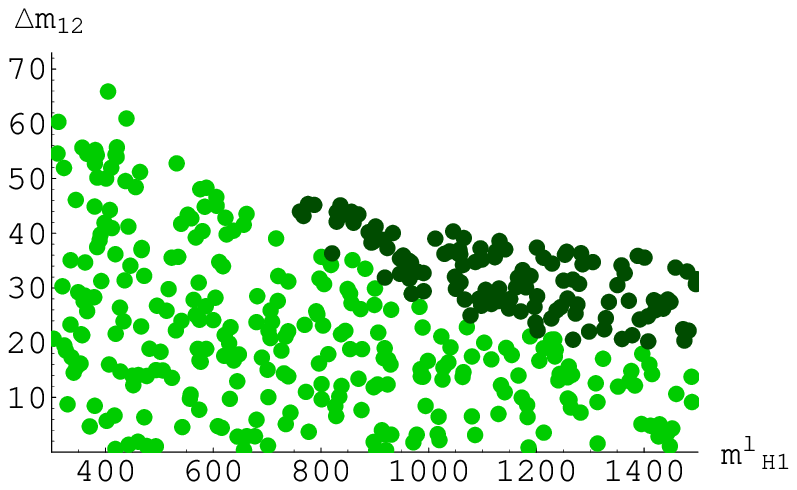,scale=.92}}\vspace{-.2cm}
\center{Scenario B}
\end{minipage}\vspace{.5cm}
\caption{\it Mirror lepton mass splitting $\Delta m_{12}=|m^\ell_{H2}-m^\ell_{H1}|$ as a function of $m^\ell_{H1}$, after imposing the existing constraints on $\mu\to e\gamma$ and $\mu^-\to e^-e^+e^-$. The dark points violate the constraint on $\mu-e$ conversion in Ti, while the light points fulfil also this constraint.\label{fig:splitting}}
\end{figure}

We also show in Fig.~\ref{fig:splitting} the mass splitting $\Delta m_{12}=|m^\ell_{H2}-m^\ell_{H1}|$ as a function of $m^\ell_{H1}$, after imposing the existing constraints on $\mu\to e\gamma$ and $\mu^-\to e^-e^+e^-$. We observe that for the scenarios A and B considered here, the first two generations of mirror leptons have to be quasi-degenerate in order to satisfy the experimental constraints. While in the case of Scenario A, $\Delta m_{12}\simle 40\gev$ is required, $\Delta m_{12}\simle 60\gev$ is sufficient to fulfil the constraints in Scenario B, provided the mirror lepton masses are relatively small. Generally we find that the allowed mass splittings are larger for smaller values of the mirror lepton masses. This is due to the fact that the functions $J^{u\bar u}$ and $J^{d\bar d}$, as defined in \eqref{Znunu} and \eqref{Zmumu}, are monotonously increasing, making thus a stronger GIM suppression necessary in the case of large masses. Finally we have studied the impact of the experimental bound on $R(\mu\text{Ti}\to e\text{Ti})$ on the allowed mass splittings. We observe that in particular for large values of the mirror lepton masses, $\mu-e$ conversion turns out to be even more constraining than $\mu\to e\gamma$ and $\mu^-\to e^-e^+e^-$.

\boldmath
\subsection{$\tau\to\mu\gamma$ and $\tau\to e\gamma$}
\unboldmath

\begin{figure}
\center{\epsfig{file=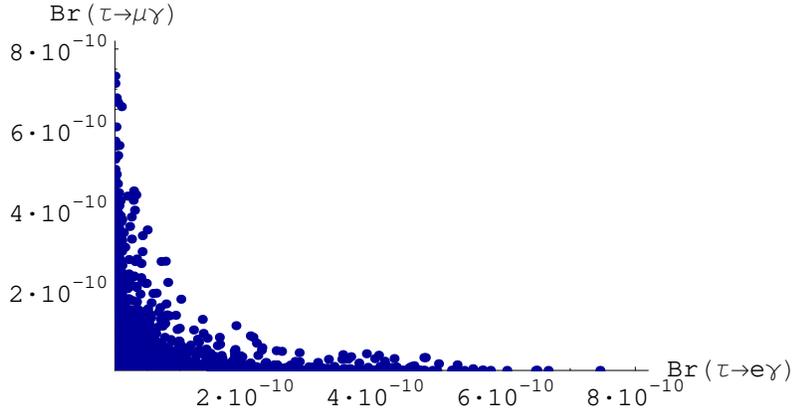,scale=1}}\vspace{-.5cm}
\caption{\it Correlation between $Br(\tau\to e\gamma)$ and $Br(\tau\to\mu\gamma)$.\label{fig:tegtmg}}
\end{figure}

In Fig.~\ref{fig:tegtmg} we show the correlation between $Br(\tau\to\mu\gamma)$ and $Br(\tau\to e\gamma)$ in the scenarios considered, imposing the experimental bounds on $\mu\to e\gamma$ and $\mu^-\to e^-e^+e^-$. We observe that they both can be individually as high as $\sim 8\cdot 10^{-10}$, but the highest values of $Br(\tau\to\mu\gamma)$ correspond generally to much lower values of $Br(\tau\to e\gamma)$ and vice versa. Still simultaneous values of both branching ratios as high as $2\cdot 10^{-10}$ are possible.

\boldmath
\subsection{$\tau\to\mu\pi,\mu\eta,\mu\eta'$ and $\tau\to\mu\gamma$}
\unboldmath

\begin{figure}
\center{\epsfig{file=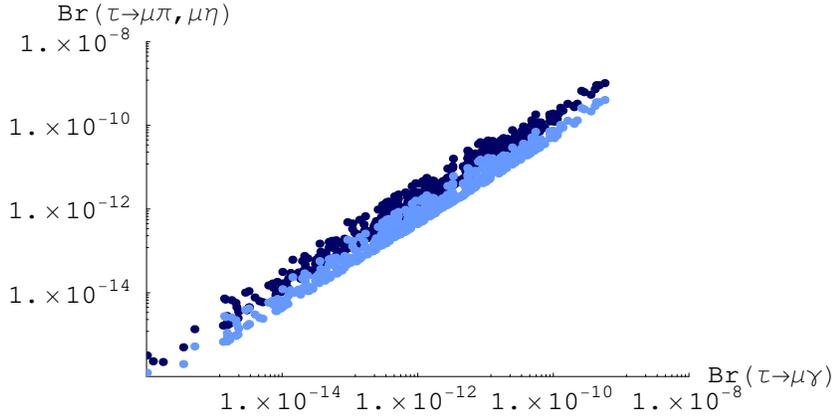,scale=1}}\vspace{-.5cm}
\caption{\it $Br(\tau\to \mu\pi)$ (dark-blue) and $Br(\tau\to \mu\eta)$ (light-blue) as functions of $Br(\tau\to\mu\gamma)$.\label{fig:tmgtmp}}
\end{figure}

In Fig.~\ref{fig:tmgtmp} we show $Br(\tau\to\mu\pi)$ and $Br(\tau\to\mu\eta)$ as functions of $Br(\tau\to\mu\gamma)$, imposing the constraints from $\mu\to e\gamma$ and $\mu^-\to e^-e^+e^-$. We find that $Br(\tau\to\mu\pi)$ can reach values as high as $2\cdot 10^{-9}$ and $Br(\tau\to\mu\eta)$ can be as large as $7\cdot 10^{-10}$, which is still by 
{more than one order of magnitude below the recent bounds from Belle and BaBar.}
We do not show $Br(\tau\to\mu\eta')$ as it is very similar to $Br(\tau\to\mu\eta)$.

Completely analogous correlations can be found also for the corresponding decays $\tau\to e\pi,e\eta,e\eta'$ and $\tau\to e\gamma$. Indeed, this symmetry between $\tau\to\mu$ and $\tau\to e$ systems turns out to be a general feature of the LHT model, that can be found in all decays considered in the present paper. We will return to this issue in Section \ref{sec:corr}.

An immediate consequence of these correlations is that, as in the case of  $\tau\to\mu\gamma$ and $\tau\to e\gamma$, the highest values for $\tau\to\mu\pi$ are possible if $\tau\to e\pi$ is relatively small, and vice versa. Still the corresponding branching ratios can be simultaneously enhanced to $3\cdot 10^{-10}$. Analogous statements apply to $\tau\to\mu(e)\eta$ and $\tau\to\mu(e)\eta'$.

\boldmath
\subsection{$K_L\to\mu e$ and $K_L\to\pi^0\mu e$}
\unboldmath

\begin{figure}
\center{\epsfig{file=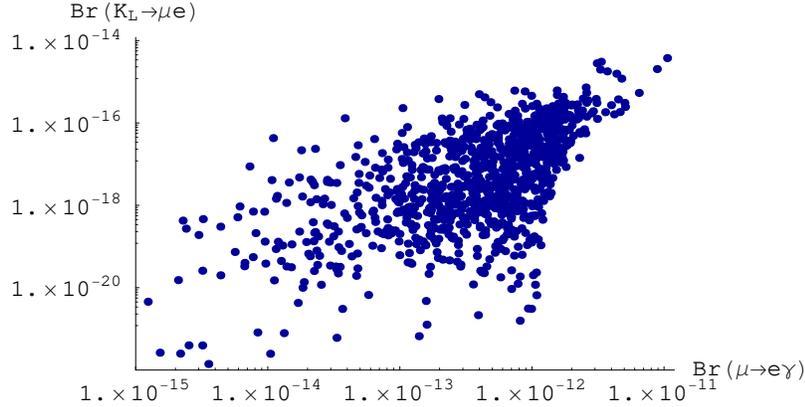,scale=0.9}}\vspace{-.4cm}
\caption{\it Correlation between $Br(\mu\to e\gamma)$ and $Br(K_L\to\mu e)$ for fixed mirror quark parameters $(m^q_{H1}\simeq332\gev,m^q_{H2}\simeq 739\gev,m^q_{H3}\simeq 819\gev,\theta^d_{12}\simeq95^\circ,\theta^d_{13}\simeq229^\circ,\theta^d_{23}\simeq205^\circ,\delta^d_{12}=0,\delta^d_{13}\simeq239^\circ,\delta^d_{23}=0)$.\label{fig:meg-KLmue}}
\end{figure}

In Fig.~\ref{fig:meg-KLmue} we show the dependence of $Br(K_L\to\mu e)$ on
$Br(\mu\to e\gamma)$. To this end we have chosen one of the sets of mirror
quark parameters for which the most spectacular effects both in $S_{\psi\phi}$
and the $K\to\pi\nu\bar\nu$ decays have been found \cite{Blanke:2006eb}. We observe
that for the parameters used here, $Br(K_L\to\mu e)$ is still by two orders of
magnitude below the current experimental upper bound. However, one can see in
Table \ref{tab:bounds} that $Br(K_L\to\mu e)$ could also be found only one
order of magnitude {below the current bound}. {This means that} large effects in the $K\to\pi\nu\bar\nu$ decays do not necessarily imply also large effects in $K_L\to\mu e$.

The effects in $K_L\to\pi^0\mu e$ are even by roughly two orders of magnitude smaller, as can also be seen in Table \ref{tab:bounds}, so that, from the point of view of the LHT model, this decay will not be observed in the foreseeable future.

\subsection{Upper Bounds}

In Table \ref{tab:bounds} we show the present LHT upper bounds on all
branching ratios considered in the present paper, together with the
corresponding experimental bounds. In order to see the strong dependence on
the scale $f$, we give these bounds both for $f=1000\gev$ and $f=500\gev$ {with the range \eqref{eq:massrange} for the mirror lepton masses in both cases.} We
observe that the upper bounds on $\tau$ decays, except for
$\tau^-\to\mu^-e^+\mu^-$ and $\tau^-\to e^-\mu^+e^-$ increase by almost two
orders of magnitude, when lowering the scale $f$ down to $500\gev$, so that
{these decays} could be found close to their current experimental upper bounds. {In particular, the most recent upper bounds on $\tau\to\mu\pi,e\pi$ \cite{Banerjee} could be violated by roughly a factor 5, as can be seen from Table \ref{tab:bounds} of the first version of the present paper. Therefore, in deriving the LHT upper bounds for $f=500\gev$, we have taken into account also the latter bounds.} On the other hand, the bounds on $\tau^-\to\mu^-e^+\mu^-$ and $\tau^-\to e^-\mu^+e^-$ are quite independent of the value of $f$. This striking difference is due to the fact that the present lepton constraints are only effective for $\mu\to e$ transitions. We also note that the upper bounds on some $K$ and $B_{d,s}$ decays, namely $K_L\to\mu e$, $K_L\to\pi^0\mu e$ and $B_{d,s}\to\mu e$, result to be lower when the NP scale is decreased to $f=500\gev$. The origin of this behaviour is that lowering $f$ the strongest constraints, mainly on $K$ and $B$ systems, start to exclude some range of parameters and, consequently, to forbid very large values for the branching ratios in question.

\begin{table}
{
\begin{center}
\begin{tabular}{|c|c|c|c|}
\hline
decay & $f=1000\gev$ & $f=500\gev$ & exp.~upper bound \\\hline\hline
$\mu\to e\gamma$ & $1.2\cdot 10^{-11}$ ($1\cdot10^{-11}$) & $1.2\cdot 10^{-11}$ ($1\cdot 10^{-11}$) & $1.2\cdot 10^{-11}$ \cite{muegamma} \\
$\mu^-\to e^-e^+e^-$ & ~$1.0\cdot 10^{-12}$ ($1\cdot 10^{-12}$)~ &~$1.0\cdot 10^{-12}$ ($1\cdot 10^{-12}$)~ & $1.0\cdot 10^{-12}$ \cite{meee} \\
$\mu\text{Ti}\to e\text{Ti}$  & $2\cdot 10^{-10}$ ($5\cdot10^{-12}$) & $4\cdot10^{-11}$ ($5\cdot10^{-12}$) & $4.3\cdot10^{-12}$ \cite{mue-conv_bound}  \\\hline
$\tau\to e\gamma$ & $8\cdot 10^{-10}$ ($7\cdot 10^{-10}$) & ${1\cdot 10^{-8}}$ (${1\cdot 10^{-8}}$) & ${9.4\cdot10^{-8}}$ \cite{Banerjee} \\
$\tau\to \mu\gamma$ & $8\cdot 10^{-10}$ ($8\cdot 10^{-10}$) &$2\cdot 10^{-8}$  (${1\cdot 10^{-8}}$) &${1.6\cdot10^{-8}}$ \cite{Banerjee}\\
$\tau^-\to e^-e^+e^-$ & $7\cdot10^{-10}$ ($6\cdot 10^{-10}$) & ${2\cdot10^{-8}}$  (${2\cdot 10^{-8}}$) & $2.0\cdot10^{-7}$ \cite{AUBERT04J}\\
$\tau^-\to \mu^-\mu^+\mu^-$ & $7\cdot10^{-10}$ ($6\cdot 10^{-10}$) & ${3\cdot10^{-8}}$  (${3\cdot 10^{-8}}$)  & $1.9\cdot10^{-7}$ \cite{AUBERT04J} \\
$\tau^-\to e^-\mu^+\mu^-$ & $5\cdot10^{-10}$ ($5\cdot 10^{-10}$)& ${2\cdot10^{-8}}$ (${2\cdot 10^{-8}}$)  & $2.0\cdot10^{-7}$ \cite{YUSA04}\\
$\tau^-\to \mu^-e^+e^-$ & $5\cdot10^{-10}$ ($5\cdot 10^{-10}$)& ${2\cdot10^{-8}}$ (${2\cdot 10^{-8}}$) &$1.9\cdot10^{-7}$ \cite{YUSA04} \\
$\tau^-\to \mu^-e^+\mu^-$ & $5\cdot10^{-14}$ ($3\cdot10^{-14}$) & ${2\cdot10^{-14}}$ (${2\cdot10^{-14}}$) & $1.3\cdot10^{-7}$ \cite{AUBERT04J}\\
$\tau^-\to e^-\mu^+e^-$ & $5\cdot10^{-14}$ ($3\cdot10^{-14}$) &${2\cdot10^{-14}}$ (${2\cdot10^{-14}}$)  & $1.1\cdot10^{-7}$ \cite{AUBERT04J} \\
$\tau\to\mu\pi$ & $2\cdot10^{-9} $ ($2\cdot10^{-9} $) & ${5.8\cdot10^{-8}}$ (${5.8\cdot10^{-8}}$) & ${5.8\cdot10^{-8}}$ \cite{Banerjee}\\
$\tau\to e\pi$ & $2\cdot10^{-9} $ ($2\cdot10^{-9} $)& ${4.4\cdot10^{-8}}$ (${4.4\cdot10^{-8}}$)   & ${4.4\cdot10^{-8}}$ \cite{Banerjee}\\
$\tau\to\mu\eta$ & $6\cdot10^{-10}$ $(6\cdot10^{-10})$ & ${2\cdot10^{-8}}$ ${(2\cdot10^{-8})}$ &  ${5.1\cdot 10^{-8}}$ \cite{Banerjee}\\
$\tau\to e\eta$ & $6\cdot10^{-10}$ $( 6\cdot10^{-10})$ & ${2\cdot10^{-8}}$ ${(2\cdot10^{-8})}$ &  ${4.5\cdot 10^{-8}}$ \cite{Banerjee}\\
$\tau\to \mu\eta'$ & $7\cdot10^{-10}$ $(7\cdot10^{-10})$& ${3\cdot10^{-8}}$ ${(3\cdot10^{-8})}$ & ${5.3\cdot 10^{-8}}$ \cite{Banerjee}\\
$\tau\to e\eta'$ & $7\cdot10^{-10}$ $(7\cdot10^{-10})$& ${3\cdot10^{-8}}$ ${(3\cdot10^{-8})}$  & ${9.0\cdot 10^{-8}}$ \cite{Banerjee}\\\hline
$K_L\to\mu e$ & $4\cdot 10^{-13}$ ($2\cdot10^{-13}$) &  $3\cdot10^{-14}$ ($3\cdot10^{-14}$)& $4.7\cdot10^{-12}$ \cite{KLmue-exp}\\
$K_L\to\pi^0\mu e$ & $4\cdot 10^{-15}$ ($2\cdot10^{-15}$)  &  $5\cdot10^{-16}$ ($5\cdot10^{-16}$) & $6.2\cdot10^{-9}$ \cite{ARISAKA98}\\
$B_d\to\mu e$ & $5\cdot10^{-16}$ ($2\cdot10^{-16}$)  & $9\cdot10^{-17}$ ($9\cdot10^{-17}$) & $1.7\cdot10^{-7}$ \cite{CHANG03}\\
$B_s\to\mu e$ & $5\cdot 10^{-15}$ ($2\cdot10^{-15}$)  &$9\cdot10^{-16}$ ($9\cdot10^{-16}$) & $6.1\cdot10^{-6}$ \cite{ABE98V}\\
$B_d\to\tau e$ & $3\cdot 10^{-11}$  ($2\cdot10^{-11}$) & ${2\cdot10^{-10}}$ ($2\cdot10^{-10}$) & $1.1\cdot10^{-4}$ \cite{BORNHEIM04}\\
$B_s\to\tau e$ & $2\cdot10^{-10}$ ($2\cdot10^{-10}$) & ${2\cdot10^{-9}}$ ($2\cdot10^{-9}$)& ---\\
$B_d\to\tau\mu$ & $3\cdot10^{-11}$ ($3\cdot10^{-11}$) & $3\cdot10^{-10}$ ($3\cdot10^{-10}$) & $3.8\cdot10^{-5}$ \cite{BORNHEIM04} \\
$B_s\to\tau\mu$ & $2\cdot10^{-10}$ ($2\cdot10^{-10}$) &  $3\cdot10^{-9}$ ($3\cdot10^{-9}$)& ---\\\hline
\end{tabular}
\end{center}
}
\caption{\it Upper bounds on LFV decay branching ratios in the LHT model, for two different values of the scale $f$, after imposing the constraints on $\mu\to e\gamma$ and $\mu^-\to e^-e^+e^-$. The numbers given in brackets are obtained after imposing the additional constraint $R(\mu\text{Ti}\to e\text{Ti})<5\cdot10^{-12}$. {For $f=500\gev$, also the bounds on $\tau\to\mu\pi,e\pi$ have been included.} The current experimental upper bounds are also given.\label{tab:bounds}}
\end{table}

We have also investigated the effect of imposing in addition the constraint $R(\mu\text{Ti}\to e\text{Ti})<5\cdot10^{-12}$, which we choose slightly above the experimental value $4.3\cdot 10^{-12}$ in order to account for the theoretical uncertainties involved. We find that all upper bounds collected in Table \ref{tab:bounds} depend only weakly on that constraint. This finding justifies that we did not take into account this bound in our numerical analysis so far, as it has only a minor impact on the observables discussed.

We would like to stress that the bounds in Table \ref{tab:bounds} should only be considered as rough upper bounds. They have been obtained from scattering over the allowed parameter space of the model. In particular, no confidence level can be assigned to them. The same applies to the ranges given in Table \ref{tab:ratios} for the LHT model.

\boldmath
\subsection{$(g-2)_\mu$}
\unboldmath

Finally, we have analyzed $(g-2)_\mu$ in the LHT model. Even for the scale $f$ being as low as $500\gev$, we find
\be
a_\mu^\text{LHT} < 1.2\cdot 10^{-10}\,,
\ee
to be compared with the experimental value in \eqref{Exp}. We observe that the effect of mirror fermions is by roughly a factor of 5 below the current experimental uncertainty, implying that the possible discrepancy between the SM value and the data cannot be solved in the model considered here.

\newsection{Patterns of Correlations and Comparison with Supersymmetry}\label{sec:corr}

\subsection{Preliminaries}

We have seen in the previous section that the branching ratios for several
charged LFV processes could reach within the LHT model the level accessible to
experiments performed in this decade. However, in view of many parameters
involved, it is desirable to look for certain correlations between various
branching ratios that are less parameter dependent than individual branching
ratios, {and whose pattern could provide a clear signature of the LHT model}.

In the case of CMFV in the quark sector {useful correlations} have been summarized at length in \cite{mfvlectures,BBGT}. In the case of LFV, in a  very interesting paper \cite{Ellis}, Ellis {\it et al.} noticed a number of correlations characteristic for the MSSM, in the absence of significant Higgs contributions. These correlations have also been analyzed recently in {\cite{Brignole,Herrero,Paradisi}}. In particular, in {\cite{Brignole,Paradisi}} modifications of them in the presence of significant Higgs contributions have been pointed out.

The main goal of this section is a brief review of the correlations discussed
in {\cite{Ellis,Brignole,Herrero,Paradisi}} and the comparison of MSSM results with the
ones of the LHT model. We will see that indeed the correlations in question
could allow for a transparent distinction between MSSM and LHT, which is
difficult in high energy collider processes \cite{mirror-phen}.

\subsection{Correlations in the MSSM}

\subsubsection{Dipole Operator Dominance}

In the absence of significant Higgs boson contributions, the LFV processes considered in our paper are dominated in the MSSM by the dipole operator with very small contributions from box and $Z$-penguin diagrams. In this case one finds the approximate general formulae {\cite{Ellis,Brignole,Herrero,Paradisi}}
\bea\label{eq:li3lj/liljg}
\frac{Br(\ell_i^-\to \ell_j^-\ell_j^+\ell_j^-)}{Br(\ell_i\to \ell_j\gamma)} &\simeq& \frac{\alpha}{3\pi}\left(\log\frac{m_{\ell_i}^2}{m_{\ell_j}^2}-2.7\right)\,,\\
\frac{Br(\ell_i^-\to \ell_j^-\ell_k^+\ell_k^-)}{Br(\ell_i\to \ell_j\gamma)} &\simeq& \frac{\alpha}{3\pi}\left(\log\frac{m_{\ell_i}^2}{m_{\ell_k}^2}-2.7\right)\,.\label{eq:lijk}
\eea
Consequently one finds
\bea\label{eq:meee/meg}
\frac{Br(\mu^-\to e^-e^+e^-)}{Br(\mu\to e\gamma)}&\simeq& 
\frac{\alpha}{3\pi}\left(\log\frac{m_{\mu}^2}{m_{e}^2}-2.7\right)\simeq\frac{1}{162} \,,\qquad\\
\frac{Br(\tau^-\to e^-e^+e^-)}{Br(\tau\to e\gamma)}\simeq
\frac{Br(\tau^-\to \mu^-e^+e^-)}{Br(\tau\to \mu\gamma)}&\simeq&\frac{\alpha}{3\pi}\left(\log\frac{m_\tau^2}{m_e^2}-2.7\right) \simeq \frac{1}{95} \,,\label{eq:teee/teg}\\
\frac{Br(\tau^-\to \mu^-\mu^+\mu^-)}{Br(\tau\to \mu\gamma)}\simeq
\frac{Br(\tau^-\to e^-\mu^+\mu^-)}{Br(\tau\to e\gamma)}&\simeq&
\frac{\alpha}{3\pi}\left(\log\frac{m_\tau^2}{m_\mu^2}-2.7\right) \simeq \frac{1}{438} \,,\label{eq:tmmm/tmg}
\eea
and \cite{Paradisi}
\be\label{eq:tmee/tmmm}
\frac{Br(\tau^-\to\mu^-e^+e^-)}{Br(\tau^-\to \mu^-\mu^+\mu^-)}\simeq
\frac{Br(\tau^-\to e^-e^+e^-)}{Br(\tau^-\to e^-\mu^+\mu^-)}\simeq 4.6\,.
\ee
Moreover, {keeping only the dipole operator contribution in $R(\mu\text{Ti}\to e\text{Ti})$, we find}
\be\label{eq:conv/meg}
\frac{R(\mu\text{Ti}\to e\text{Ti})}{Br(\mu\to e\gamma)}\simeq 0.7\alpha\,.
\ee

One also has
\be\label{eq:Pgamma}
\frac{Br(\tau\to\ell P)}{Br(\tau\to\ell\gamma)} < \ord(\alpha)\qquad (P=\pi,\eta,\eta')\,,
\ee
where the absence of dipole operator contributions to $Br(\tau\to\ell P)$ makes the ratios in \eqref{eq:Pgamma} significantly smaller than the ones in \eqref{eq:meee/meg}--\eqref{eq:tmmm/tmg}.

\subsubsection{Including Higgs Contributions: Decoupling Limit}

 It should be emphasized that the Higgs contributions become competitive with the gauge mediated ones, once the Higgs masses are roughly by one order of magnitude smaller than the sfermion masses and $\tan\beta$ is $\ord(40-50)$. There is a rich literature on Higgs contributions to LFV within supersymmetry. One of the earlier references is the one by Babu and Kolda \cite{BK}. Here we concentrate on the modifications of the correlations discussed above in the presence of significant Higgs contributions as analyzed by Paradisi \cite{Paradisi}, where further references can be found {(see in particular \cite{Brignole}).}

{In the limit of  Higgs decoupling}, {which is achieved in the MSSM,} some of the results in \eqref{eq:li3lj/liljg}--\eqref{eq:Pgamma} are modified. In particular
\be
\frac{Br(\tau^-\to\mu^-\mu^+\mu^-)}{Br(\tau\to\mu\gamma)} \simle \frac{2}{9}\,,\qquad
\frac{Br(\tau^-\to e^-\mu^+\mu^-)}{Br(\tau\to e\gamma)} \simle\frac{1}{12}
\ee
can be much larger than in the case of dipole operator dominance, and
\be\label{eq:tmee/tmmm-decoupl}
\frac{Br(\tau^-\to\mu^-e^+e^-)}{Br(\tau^-\to\mu^-\mu^+\mu^-)} \simge 0.05\,,\qquad
\frac{Br(\tau^-\to e^-e^+e^-)}{Br(\tau^-\to e^-\mu^+\mu^-)} \simge 0.13
\ee
can be much smaller. 
Moreover one finds
\be\label{eq:eta-ratios}
\frac{Br(\tau\to\ell\eta)}{Br(\tau\to\ell\gamma)}\simle 1\,,\qquad
\frac{Br(\tau\to\mu\eta)}{Br(\tau^-\to\mu^-\mu^+\mu^-)} \simeq 4.5\,,\qquad
\frac{Br(\tau\to e\eta)}{Br(\tau^-\to e^-\mu^+\mu^-)} \simeq 12\,. 
\ee

\subsection{Correlations in the LHT Model}

\begin{figure}
\center{\epsfig{file=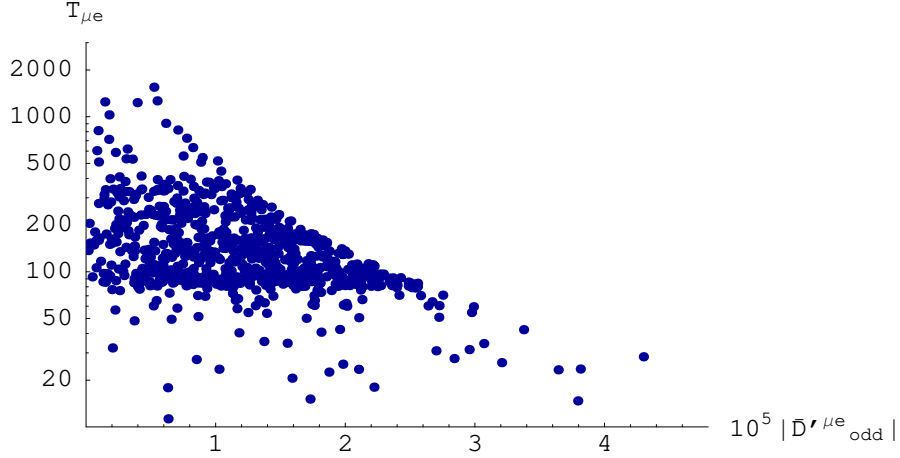,scale=1}}\vspace{-.5cm}
\caption{\it $T_{\mu e}$ as a function of $|\bar D^{\prime\,\mu e}_\text{odd}|$ for $f=1\tev$.\label{fig:Tme}}
\end{figure}

The pattern of correlations in the LHT model differs significantly {from the
MSSM one} presented above. This is due to the fact that the dipole contributions to the decays $\ell_i^-\to \ell_j^-\ell_j^+\ell_j^-$ and $\ell_i^-\to \ell_j^-\ell_k^+\ell_k^-$ can be fully neglected in comparison with $Z^0$-penguin and box diagram contributions. This is dominantly due to the fact that the neutral gauge boson $(Z_H,A_H)$ contributions interfere destructively with the $W_H^\pm$ contributions to the dipole operator functions $\bar D^{\prime\,ij}_\text{odd}$, but constructively in the case of the functions $\bar Y^{ij}_{k,\text{odd}}$ that summarize the $Z^0$-penguin and box contributions in a gauge independent manner. Moreover, the large $\tan\beta$ enhancement of dipole operators characteristic for the MSSM is absent here. In this context it is useful to define the ratios 
\be
T_{ij}=\left| \frac{\bar Y^{ij}_{j,\text{odd}}}{ \bar D^{\prime\,ij}_\text{odd}}\right|^2
\ee
that are strongly enhanced in the case of the LHT model for almost the entire space of parameters, as seen in Fig.~\ref{fig:Tme} for the case of $T_{\mu e}$.
Consequently, the logarithmic enhancement of dipole contributions seen in \eqref{eq:Brmeee}, \eqref{eq:li3lj/liljg} and \eqref{eq:lijk} is eliminated by the strong suppression of $|\bar D^{\prime\,ij}_\text{odd}|^2$ with respect to $|\bar Y^{ij}_{j,\text{odd}}|^2$. Similarly, $C_7^{\tau\mu}$, that is governed by $\bar D^{\prime\,\tau\mu}_\text{odd}$, can be neglected in \eqref{eq:Rs}, resulting in the absence of large logarithms in the decays $\tau^-\to\mu^-e^+e^-$ and $\tau^-\to e^-\mu^+\mu^-$.

{We note that the ratios $T_{ij}$ depend on the approximation made for the left-over singularity, as $\bar Y^{ij}_{j,\text{odd}}$ suffers from  this divergence while $\bar D^{'\,ij}_\text{odd}$ does not. However, we find that dropping completely the term proportional to $S_\text{odd}$, which is certainly not a good approximation, amounts to at most 50\% changes in the allowed ranges for $T_{ij}$ and consequently for the correlations discussed below. Therefore we think that these correlations are only insignificantly affected by this general feature of non-linear sigma models.} 

These findings make clear that the correlations between various branching
ratios in the LHT model, being unaffected by the logarithms present in
\eqref{eq:li3lj/liljg}--\eqref{eq:tmmm/tmg}, {have a very different pattern from those found in the MSSM}.

It turns out that within an accuracy of $3\%$ one can set $\bar D^{\prime\,ij}_\text{odd}$ to zero in all decays with three leptons in the final state. Similarly, the contribution of $\tilde C_9^{ij}$ in $\tau^-\to\mu^-e^+e^-$ and $\tau^-\to e^-\mu^+\mu^-$ can be neglected. Neglecting finally $\Delta_{ij}$ in \eqref{eq:C9}, which is good to within 20\%, allows us to derive simple expressions for various ratios of branching ratios that can directly be compared with those listed in \eqref{eq:li3lj/liljg}--\eqref{eq:tmee/tmmm}.

To this end, we define
\begin{gather}
a_{ij} = \frac{\bar Z^{ij}_\text{odd}}{\bar Y^{ij}_{j,\text{odd}}}\,,\qquad
b^k_{ij} = \left|\frac{\bar Y^{ij}_{k,\text{odd}}}{\bar Y^{ij}_{j,\text{odd}}}\right|^2\,,\\
c_{ij} = 3\sin^4\theta_W |a_{ij}|^2 +\frac{1}{2}-2\sin^2\theta_W\text{Re}{(a_{ij})}\,.
\end{gather}

We find then
\be\label{eq:meee/meg-LHT}
\frac{Br(\mu^-\to e^-e^+e^-)}{Br(\mu\to e\gamma)}\simeq
\frac{2\alpha}{3\pi}\frac{1}{\sin^4\theta_W}T_{\mu e}c_{\mu e}\,,
\ee
with analogous expressions for the respective $\tau\to e$ and $\tau\to\mu$ transitions.

We also find
\bea
\frac{Br(\tau^-\to e^-\mu^+\mu^-)}{Br(\tau\to e\gamma)}&\simeq& \frac{\alpha}{12\pi}\frac{1}{\sin^4\theta_W}T_{\tau e}b^\mu_{\tau e}\,.\label{eq:temm/teg-LHT}\\
\frac{Br(\tau^-\to\mu^-e^+e^-)}{Br(\tau\to\mu\gamma)}&\simeq& \frac{\alpha}{12\pi}\frac{1}{\sin^4\theta_W}T_{\tau\mu}b^e_{\tau\mu}\,,\label{eq:tmee/tmg-LHT}
\eea
Finally, \eqref{eq:meee/meg-LHT}--\eqref{eq:tmee/tmg-LHT} imply
\bea
\frac{Br(\tau^-\to e^-e^+e^-)}{Br(\tau^-\to e^-\mu^+\mu^-)} &=& 8\frac{c_{\tau e}}{b^\mu_{\tau e}} \,,\label{eq:teee/temm-LHT}\\
\frac{Br(\tau^-\to \mu^-\mu^+\mu^-)}{Br(\tau^-\to \mu^-e^+e^-)} &=& 8\frac{c_{\tau \mu}}{b^e_{\tau \mu}} \,.\label{eq:tmmm/tmee-LHT}
\eea

{In what follows we restrict ourselves to $f=1\tev$ for simplicity. We note that although the numerical values given below depend slightly on the size of the scale $f$, the qualitative picture found and discussed remains true independently of that value.}
   
The ranges for the ratios in question found in the LHT model are compared in Table \ref{tab:ratios} with the corresponding values in the MSSM, both in the case of dipole dominance and when Higgs contributions are significant.

\begin{table}
{\renewcommand{\arraystretch}{1.5}
\begin{center}
\begin{tabular}{|c|c|c|c|}
\hline
ratio & LHT  & MSSM (dipole) & MSSM (Higgs) \\\hline\hline
$\frac{Br(\mu^-\to e^-e^+e^-)}{Br(\mu\to e\gamma)}$  & \hspace{.8cm} 0.4\dots2.5\hspace{.8cm}  & $\sim6\cdot10^{-3}$ &$\sim6\cdot10^{-3}$  \\
$\frac{Br(\tau^-\to e^-e^+e^-)}{Br(\tau\to e\gamma)}$   & 0.4\dots2.3     &$\sim1\cdot10^{-2}$ & ${\sim1\cdot10^{-2}}$\\
$\frac{Br(\tau^-\to \mu^-\mu^+\mu^-)}{Br(\tau\to \mu\gamma)}$  &0.4\dots2.3     &$\sim2\cdot10^{-3}$ & $0.06\dots0.1$ \\
$\frac{Br(\tau^-\to e^-\mu^+\mu^-)}{Br(\tau\to e\gamma)}$  & 0.3\dots1.6     &$\sim2\cdot10^{-3}$ & $0.02\dots0.04$ \\
$\frac{Br(\tau^-\to \mu^-e^+e^-)}{Br(\tau\to \mu\gamma)}$  & 0.3\dots1.6    &$\sim1\cdot10^{-2}$ & ${\sim1\cdot10^{-2}}$\\
$\frac{Br(\tau^-\to e^-e^+e^-)}{Br(\tau^-\to e^-\mu^+\mu^-)}$     & 1.3\dots1.7   &$\sim5$ & 0.3\dots0.5\\
$\frac{Br(\tau^-\to \mu^-\mu^+\mu^-)}{Br(\tau^-\to \mu^-e^+e^-)}$   & 1.2\dots1.6    &$\sim0.2$ & 5\dots10 \\
$\frac{R(\mu\text{Ti}\to e\text{Ti})}{Br(\mu\to e\gamma)}$  & $10^{-2}\dots 10^2$     & $\sim 5\cdot 10^{-3}$ & $0.08\dots0.15$ \\\hline
\end{tabular}
\end{center}\renewcommand{\arraystretch}{1.0}
}
\caption{\it Comparison of various ratios of branching ratios in the LHT model and in the MSSM without and with significant Higgs contributions.\label{tab:ratios}}
\end{table}

While the results in Table \ref{tab:ratios} are self-explanatory, let us
emphasize four striking differences {between the LHT and MSSM results} in the case of small Higgs contributions:
\bi
\item
The ratio \eqref{eq:meee/meg-LHT} and {the similar ratios} for $\tau\to e$ and $\tau\to\mu$ transitions are $\ord(1)$ in the LHT model as opposed to $\ord(\alpha)$ in the MSSM.
\item {Also the $\mu-e$ conversion rate in nuclei, normalized to $Br(\mu\to e\gamma)$, can be significantly enhanced in the LHT model, with respect to the MSSM without significant Higgs contributions. However the distinction in this case is not as clear as in the case of $Br(\ell_i^-\to\ell_j^-\ell_j^+\ell_j^-)/Br(\ell_i\to\ell_j\gamma)$ due to a destructive interference of the different contributions to $\mu-e$ conversion (see \eqref{eq:mueconv}). On the other hand, in the case of the MSSM with significant Higgs contributions, $R(\mu\text{Ti}\to e\text{Ti})/Br(\mu\to e\gamma)$ is typically much larger than $\alpha$, so that a distinction from the LHT model would be difficult in that case. }
\item
The ``inverted'' pattern of the ratios in \eqref{eq:tmee/tmmm} is absent in the LHT model as the ratios \eqref{eq:teee/temm-LHT} and \eqref{eq:tmmm/tmee-LHT} are comparable in magnitude. Moreover they are generally very different from the MSSM values.
\item
The last finding also implies that while the ratios \eqref{eq:teee/teg} and \eqref{eq:tmmm/tmg} differ roughly by a factor of 5 in the case of the MSSM, they are comparable in the LHT model, as seen in  \eqref{eq:temm/teg-LHT} and \eqref{eq:tmee/tmg-LHT}.
\ei

In order to exhibit these different patterns in a transparent manner, let us define the three ratios
\bea
R_1 &=& \frac{Br(\tau^-\to e^-e^+e^-)}{Br(\tau^-\to \mu^-\mu^+\mu^-)} \frac{Br(\tau^-\to \mu^-e^+e^-)}{Br(\tau^-\to e^-\mu^+\mu^-)}\,, \\
R_2 &=& \frac{Br(\tau^-\to e^-e^+e^-)}{Br(\tau^-\to \mu^-\mu^+\mu^-)} \frac{Br(\tau\to \mu\gamma)}{Br(\tau\to e\gamma)}\,,\\
R_3 &=& \frac{Br(\tau^-\to e^-\mu^+\mu^-)}{Br(\tau^-\to \mu^-e^+e^-)}  \frac{Br(\tau\to \mu\gamma)}{Br(\tau\to e\gamma)}\,.
\eea
Note that in the case of a $\mu\leftrightarrow e$ symmetry, these three ratios
should be equal to unity. This symmetry is clearly very strongly broken in the
MSSM, due to the sensitivity of the ratios in \eqref{eq:li3lj/liljg} and
\eqref{eq:lijk} to $m_e$ and $m_\mu$, {where one finds} 
\be\label{eq:Ri-MSSM}
R_1\simeq {20}\,,\qquad R_2\simeq {5}\,,\qquad R_3\simeq0.2\qquad\text{(MSSM)}\,.
\ee
On the other hand, in the case of the LHT model the absence of large logarithms allows to satisfy the $\mu\leftrightarrow e$ symmetry in question within 30\%, so that we find
\be
0.8\simle R_1 \simle1.3\,,\qquad
0.8\simle R_2 \simle1.2\,,\qquad
0.8\simle R_3 \simle1.2\qquad\text{(LHT)}\,.\label{eq:Ri-LHT}
\ee
The comparison of \eqref{eq:Ri-MSSM} with \eqref{eq:Ri-LHT} offers a very clear distinction between these two models.

In the presence of significant Higgs contributions the distinction between MSSM and LHT is less pronounced in $\tau$ decays with $\mu$ in the final state. However even in this case both models can be distinguished, as seen in Table \ref{tab:ratios}. In particular, the four ratios involving $Br(\ell_i\to\ell_j\gamma)$ are still significantly smaller in the MSSM than in the LHT model, and all decays with electrons in the final state offer excellent means to distinguish these two models.

For the decays $\tau\to\ell P$  with $P=\pi,\eta,\eta'$ we find the ranges
\be
1\simle \frac{Br(\tau\to\ell\pi)}{Br(\tau\to\ell\gamma)}\simle5.5\,,\qquad
0.4\simle \frac{Br(\tau\to\ell\eta)}{Br(\tau\to\ell\gamma)}\simle2\,,\qquad
0.3\simle \frac{Br(\tau\to\ell\eta')}{Br(\tau\to\ell\gamma)}\simle2.8\,.
\ee
As seen in \eqref{eq:Pgamma}, the ratios obtained in the LHT model are much larger than in supersymmetry without significant Higgs contributions. On the other hand, if the Higgs contributions are dominant, the distinction through the first inequality of \eqref{eq:eta-ratios} will be difficult. However, we also find
\be
0.7\simle\frac{Br(\tau\to\mu\eta)}{Br(\tau^-\to\mu^-\mu^+\mu^-)}\simle 1.3\,,\qquad 
1.1\simle\frac{Br(\tau\to e\eta)}{Br(\tau^-\to e^-\mu^+\mu^-)}\simle 1.8\,,
\ee
that could be distinguished from the corresponding results {in \eqref{eq:eta-ratios}}.

In summary we have demonstrated that the LHT model can be very transparently distinguished from the MSSM with the help of LFV processes, while such a distinction is non-trivial in the case of high energy processes \cite{mirror-phen}. We consider this result as one of the most interesting ones of our paper.

\newsection{Conclusions}\label{sec:concl}

In the present paper we have extended our analysis of flavour and CP-violating processes in the LHT model \cite{BBPTUW,Blanke:2006eb} to the lepton sector.

In contrast to rare $K$ and $B$ decays, where the SM contributions play an important and often the dominant role in the LHT model, the smallness of ordinary neutrino masses assures that the mirror fermion contributions to LFV processes are by far the dominant effects. Moreover, the absence of QCD corrections and hadronic matrix elements allows in most cases to make predictions entirely within perturbation theory. Exceptions are the decays $B_{d,s}\to\ell_i\ell_j$ that involve the weak decay constants $F_{B_{d,s}}$.

The decays and transitions considered by us can be divided into two broad classes: those which suffer from some sensitivity to the UV completion signalled by the logarithmic dependence on the cut-off (class A) and those which are free from this dependence (class B). We have
\vspace{.2cm}

\underline{Class A}
\begin{gather*}
\mu^-\to e^-e^+e^-\,,\qquad \tau^-\to e^-e^+e^-\,,\qquad \tau^-\to \mu^-\mu^+\mu^-\,,\qquad \mu\to e \text{ conversion}\,,\\
\tau^-\to \mu^-e^+e^-\,,\qquad \tau^-\to e^-\mu^+\mu^-\,,\qquad
{\tau\to\mu (e) \pi\,,\qquad \tau\to \mu (e) \eta\,,\qquad \tau\to\mu (e) \eta'}\,.
\end{gather*}

\underline{Class B}
\begin{gather*}
\mu\to e\gamma\,,\qquad \tau\to e\gamma\,,\qquad \tau\to\mu\gamma\,,\\
K_{L,S}\to\mu e\,,\qquad K_{L,S}\to\pi^0\mu e\,,\qquad B_{d,s}\to\mu e\,,\qquad B_{d,s}\to\tau e\qquad B_{d,s}\to\tau\mu\,,\\
\tau^-\to e^-\mu^+e^-\,,\qquad \tau^-\to\mu^- e^+\mu^-\,,\qquad (g-2)_\mu\,.
\end{gather*}

Clearly the predictions for the decays in class B are more reliable, but we believe that also our estimates of the rates of class A decays give at least correct orders of magnitude. Moreover, as pointed out in \cite{Blanke:2006eb}, the logarithmic divergence in question has a universal character and can simply be parameterized by a single parameter $\delta_\text{div}$ that one can in principle fit to the data and trade for one observable. At present this is not feasible, but could become realistic when more data for FCNC processes will be available. This reasoning assumes that   $\delta_\text{div}$ encloses all effects coming from the UV completion, which is true if light fermions do not have a more complex relation to the fundamental fermions of the UV completion, that could spoil its flavour independence.

Bearing this in mind the main messages of our paper are as follows:
\bi
\item
As seen in Table~\ref{tab:bounds}, several rates considered by us can reach the present experimental upper bounds, and are very interesting in view of new experiments taking place in this decade. In fact, in order to suppress the $\mu\to e \gamma$ and $\mu^-\to e^-e^+e^-$ decay rates and the $\mu-e$ conversion rate in nuclei below the present experimental upper bounds, the relevant mixing matrix in the mirror lepton sector, $V_{H\ell}$, must be rather hierarchical, unless the spectrum of mirror leptons is quasi-degenerate.
\item
The correlations between various branching ratios analyzed in detail in
Section \ref{sec:corr} should allow a clear distinction of the LHT model from
the MSSM. While in the MSSM {without significant Higgs contributions} the dominant role in decays with three leptons in
the final state and in $\mu-e$ conversion in nuclei is played by the dipole
operator, this operator is basically irrelevant in the LHT model, where
$Z^0$-penguin and box diagram contributions are {much more important}.
{While in the Higgs mediated case, the distinction of the MSSM from the LHT model is less pronounced, all ratios involving $\ell_i\to\ell_j\gamma$ and in particular decays with electrons in the final state still offer excellent means to distinguish these two models.}
\item
The measurements of all rates considered in the present paper should allow the full determination of the matrix $V_{H\ell}$, provided the masses of the mirror fermions and of the new heavy gauge bosons will be measured at the LHC.
\item 
{We point out that the measurements of $Br(K_L\to\mu e)$ and $Br(K_L\to\pi^0\mu e)$ will transparently shed some light on the complex phases present in the mirror quark sector.}
\item
The contribution of mirror leptons to $(g-2)_\mu$ is negligible. {This should also be contrasted with the MSSM with large $\tan\beta$ and not too heavy scalars, where those corrections could be significant, thus allowing to solve the possible discrepancy between SM prediction and experimental data \cite{EWg-2}.}
\item
{Another possibility to 
{distinguish different NP models} 
through LFV processes is given by the measurement of $\mu\to e\gamma$ with polarized muons. Measuring the angular distribution of the outgoing electrons, one can determine the size of left- and right-handed contributions separately \cite{polarized-meg}. In addition, detecting also the electron spin would yield information on the relative phase between these two contributions \cite{Farzan}. We recall that the LHT model is peculiar in this respect as it does not involve any right-handed contribution.}
\item
It will be interesting to watch the experimental progress in LFV in the coming years with the hope to see some spectacular effects of mirror fermions in LFV decays that in the SM are basically unmeasurable. The correlations between various branching ratios analyzed in detail in Section \ref{sec:corr} should be very useful in distinguishing the LHT model from other models, in particular the MSSM. In fact, this distinction should be easier than through high-energy processes at LHC, as LFV processes are theoretically very clean.
\ei

The decays $\mu\to e\gamma$, $\tau\to\mu\pi$ and $(g-2)_\mu$ have already been
analyzed in the LHT model in \cite{Goyal,IndianLFV}. While we agree with these
papers that {mirror lepton effects} in   $\mu\to e\gamma$ and $\tau\to\mu\pi$ can be very large and are very small in $(g-2)_\mu$, we disagree at the quantitative level, as discussed in the text.

\subsection*{Acknowledgements}

Our particular thanks go to Stefan Recksiegel for providing us the mirror
quark parameters necessary for the study of the $K$ and $B_{d,s}$ decays, and
to Paride Paradisi for very informative {discussions on his work.}
We would also like to thank Wolfgang Altmannshofer, Gerhard Buchalla, Vincenzo Cirigliano, Andrzej Czarnecki and Michael Wick for useful discussions. This
research was partially supported by {the Cluster of Excellence `Origin and Structure of the Universe' and by} the German `Bundesministerium f{\"u}r Bildung und
Forschung' under contract 05HT6WOA.

\begin{appendix}

\newsection{Neutrino Masses in the LHT Model}\label{sec:app1}

Within the LH model without T-parity, there have been several suggestions how to naturally explain the smallness of neutrino masses within the low-energy framework of the model \cite{LNV1,LNV2,LNV3,LNV4}. Although differing in the details, they are all based on a coupling of the form
\be\label{eq:Phicoupl}
Y_{ij} (L^i)^T \Phi \mathcal{C}^{-1} L^j +h.c.\,,
\ee
where $L^i$ are the left-handed SM lepton doublets and $\mathcal{C}$ is the charge conjugation operator. This term violates lepton number by $\Delta L=2$ and generates Majorana masses for the left-handed neutrinos of size
\be
m_{ij}=Y_{ij} v'\,,
\ee
with $v'$ being the VEV of the scalar triplet $\Phi$.

Thus, in order to explain the observed smallness of neutrino masses, either $Y_{ij}$ or $v'/v$ has to be of $\ord(10^{-11})$. While in the case of $Y_{ij}$, this appears to be an extremely fine-tuned scenario, in the case of $v'$ some so far unknown mechanism could be at work that ensures the smallness of $v'$. Such a mechanism would also be very welcome from the point of view of electroweak precision observables.

Indeed, \eqref{eq:Phicoupl} is a concrete example for the general mechanism found and discussed in \cite{tripletgen} to generate Majorana neutrino masses for the left-handed neutrinos through their interaction with a triplet scalar field.

One should however be aware of the fact that  \eqref{eq:Phicoupl} explicitly breaks the enlarged $[SU(2)\times U(1)]^2$ gauge symmetry of the LH model, while it is invariant under $SU(2)_L\times U(1)_Y$. Consequently, in \cite{LNV3} the interaction \eqref{eq:Phicoupl} has been encoded in the gauge-invariant expression
\be\label{eq:Sigmacoupl}
Y_{ij}(L^i)^T \Sigma^*\mathcal{C}^{-1} L^j +h.c.
\,,
\ee
where $\Sigma$ is an $SU(5)$ symmetric tensor containing, amongst others, the scalar triplet $\Phi$.
Now, $m_{ij}$ is given by
\be\label{eq:MMM}
m_{ij}=Y_{ij}\left(v'+\frac{v^2}{4f}\right)\,,
\ee
so that $Y_{ij}\sim \ord(10^{-11})$ is necessarily required in order to suppress also the second contribution.

In the case of the LHT model, an interaction of the form \eqref{eq:Phicoupl} is forbidden by T-parity. However, one could T-symmetrize the interaction term \eqref{eq:Sigmacoupl}, leading to
\be
Y_{ij}\bigg[(L^i_1)^T \Sigma^*\mathcal{C}^{-1} L^j_1 + ( L^i_2)^T \Omega \Sigma \Omega\mathcal{C}^{-1} L^j_2\bigg] +h.c.\,,
\ee
where $\Omega= \text{diag}(1,1,-1,1,1)$.
{In this way,} a neutrino mass matrix
\be
m_{ij}=Y_{ij}\frac{v^2}{4f}
\ee
is generated. Note that this corresponds to only the second term in 
\eqref{eq:MMM}, as the first term is forbidden by T-parity. So again we are forced to fine-tune $Y_{ij}$ to be $\ord(10^{-11})$, which is almost as unnatural as just introducing a standard Yukawa coupling to make the neutrinos massive.

A different way to implement naturally small neutrino masses in Little Higgs
models has been {developed} for the Simplest Little Higgs model in \cite{Lee:SHneutrino,Aguila}. Here, three TeV-scale Dirac neutrinos have been introduced, with a small ($\sim 0.1\kev$) lepton number violating Majorana mass term. Like that, naturally small masses for the SM neutrinos are generated radiatively. This idea can easily be applied to the LHT model. However, as already discussed in \cite{Aguila}, the mixing of the SM neutrinos with the heavy Dirac neutrinos appears at $\ord(v/f)$, so that $f$ is experimentally constrained in this framework to be at least $\sim 3-4\tev$. This bound is much stronger than the one coming from electroweak precision constraints, $f\simge 500\gev$ \cite{mH}, and re-introduces a significant amount of fine-tuning in the theory. For that reason, we do not follow this approach any further. 

Thus so far we did not find a satisfactory way to naturally explain the smallness of neutrino masses in the LHT model. In fact, it is easy to understand why this does not work. In order to keep the relevant couplings of $\ord(1)$, there should be either a very small scale, as $v'$ in the LH model, or a large hierarchy of scales, as in the see-saw mechanism \cite{seesaw}, present in the theory\footnote{Note that also the smallness of the lepton number violating scale in the model of \cite{Aguila} would require some explanation.}. However, in the LHT model the only relevant scales are
\be
v=246\gev\,,\qquad f\sim 1\tev\,,\qquad 4\pi f\sim 10\tev\,,
\ee
which are all neither small enough nor widely separated.

To conclude, we find that the LHT model cannot naturally explain the observed smallness of neutrino masses. This, in our opinion, should however not be understood as a failing of the model. After all, the LHT model is an effective theory with a cutoff of $\ord(10\tev)$, while the generation of neutrino masses is, as in see-saw models, usually understood to  be related to some much higher scale, which will in turn be described by the UV completion of the model. Consequently, in our analysis we have simply assumed that the mechanism for generating neutrino masses is incorporated in the (unspecified) UV completion, and that the details of this mechanism have only negligible impact on the low-energy observables studied in the present paper.

\newsection{Relevant Functions}\label{sec:app2}

\bea
D_0(x)&=&  -\frac{4}{9}\log x+\frac{-19x^3+25x^2}{36(x-1)^3}+\frac{x^2(5x^2-2x-6)}{18(x-1)^4}\log x\,,\\
E_0(x)&=&  -\frac{2}{3}\log{x}+\frac{x^2(15-16x+4x^2)}{6(1-x)^4}\log{x} + \frac{x(18-11x-x^2)}{12(1-x)^3}\,,\\
 D'_0(x)&=&-\dfrac{3x^3-2x^2}{2(x-1)^4}\log x +
\dfrac{8x^3+5x^2-7x}{12(x-1)^3}\,,\\
E'_0(x)&=&\dfrac{3x^2}{2(x-1)^4}\log x +
\dfrac{x^3-5x^2-2x}{4(x-1)^3}\,.
\eea

\bea
R_2(y_i) &=& -\left[\frac{y_i \log y_i}{(1-y_i)^2}+\frac{1}{1-y_i}\right]\,,\\
F_2(y_i) &=& -\frac{1}{2}\left[\frac{y_i^2 \log
    y_i}{(1-y_i)^2}+\frac{1}{1-y_i}\right]\,.
\eea

\begin{eqnarray}
F^{u \bar{u}}\left(y_{i}, z; W_{H}\right) &=& \frac{3}{2} y_{i} - F_{5}\left(y_{i}, z\right) - 7 F_{6}\left(y_{i}, z\right) - 9 U\left(y_{i}, z\right)\,, \\
F^{d\bar d}\left(y_{i}, z; W_{H}\right) &=& \frac{3}{2} y_{i} - F_{5}\left(y_{i}, z\right)- 7 F_{6}\left(y_{i}, z\right) + 3 U\left(y_{i}, z\right)\,.
\end{eqnarray}

\begin{eqnarray}
F_{5}\left(y_{i}, z\right) &=& \frac{y_{i}^{3} \log y_{i}}{\left(1-y_{i}\right) \left(z-y_{i}\right)} + \frac{z^{3} \log z}{\left(1-z\right) \left(y_{i}-z\right)}\,, \\
F_{6}\left(y_{i}, z\right) &=& -\left[\frac{y_{i}^{2} \log y_{i}}{\left(1-y_{i}\right) \left(z-y_{i}\right)} + \frac{z^{2} \log z}{\left(1-z\right) \left(y_{i}-z\right)}\right]\,, \\
U\left(y_{i}, z\right) &=& \frac{y_{i}^{2} \log y_{i}}{\left(y_{i}-z\right) \left(1-y_{i}\right)^{2}} + \frac{z^{2} \log z}{\left(z-y_{i}\right) \left(1-z\right)^{2}} + \frac{1}{\left(1-y_{i}\right) \left(1-z\right)}\,.
\end{eqnarray}

\begin{eqnarray}
G\left(y_{i}, z; Z_{H}\right) &=&  -\frac{3}{4} U\left(y_i, z\right) \,,\\
G_{1}\left(y_{i}^{\prime},z^{\prime}; A_{H}\right) &=& \frac{1}{25 a} G\left(y_{i}^{\prime}, z^{\prime}; Z_{H}\right) \,,\\
G_{2}\left(y_{i}, z; \eta\right) &=& -\frac{3}{10 a} \left[\frac{y_{i}^{2} \log y_{i}}{\left(1-y_{i}\right) \left(\eta-y_{i}\right) \left(y_{i}-z\right)} \right.\nn \\
&& +  \left. \frac{z^{2} \log z}{\left(1-z\right) \left(\eta-z\right) \left(z-y_{i}\right)} + \frac{\eta^{2} \log \eta}{\left(1-\eta\right) \left(y_{i}-\eta\right) \left(\eta-z\right)}\right]\,.
\end{eqnarray}

\bea
C_\text{odd}(y_i)&=& \frac{1}{64}\frac{v^2}{f^2}\left[y_i S_\text{odd}-8 y_i R_2(y_i) + \frac{3}{2}y_i+2y_iF_2(y_i)\right],   \\
D_\text{odd}(y_i)&=&   \frac{1}{4}\frac{v^2}{f^2}\left[D_0(y_i)-\frac{7}{6}E_0(y_i)-\frac{1}{10}{E_0}{(y'_i)} \right],\\
S_\text{odd}&=&\frac{1}{\varepsilon}+\log\frac{\mu^2}{M_{W_H}^2} \longrightarrow \log\frac{(4\pi f)^2}{M_{W_H}^2}\,.
\eea

\begin{eqnarray}
F(z_i,y_j;W_H)&=& \frac{1}{(1-z_i)(1-y_j)} \left(1-\frac{7}{4} z_i
y_j\right) +\frac{z_i^2 \log z_i}{(z_i - y_j) (1-z_i)^2} \left( 1- 2
y_j + \frac{z_i y_j}{4} \right)\nn\\
&& -\frac{y_j^2 \log y_j}{(z_i - y_j) (1-y_j)^2} \left( 1- 2
z_i + \frac{z_i y_j}{4} \right), \\\nn\\
A_1 (z_i, y_j ; Z_H ) &=& -\frac{3}{100 a} \left[ \frac{1}{(1- z'_i)(1-y'_j)} + \frac{z'_i z_i \log
    z'_i}{(z_i-y_j) (1-z'_i)^2}\right. \nn\\
&&\left.  - \frac{y'_j y_j \log
    y'_j}{(z_i-y_j) (1-y'_j)^2} \right], \\\nn\\
A_2 (z_i, y_j ; Z_H ) &=& -\frac{3}{10} \left[ \frac{\log a}{(a-1) (1- z'_i)(1-y'_j)} + \frac{z_i^2 \log
    z_i}{(z_i-y_j) (1-z_i) (1- z'_i)}\right. \nn\\
&&\left.  - \frac{y_j^2 \log
    y_j}{(z_i-y_j) (1-y_j)(1-y'_j)} \right].
\end{eqnarray}

\bea
L_1(y_i)&=&\frac{1}{12(1-y_i)^4}\left[-8+38y_i-39y_i^2+14y_i^3-5y_i^4+18y_i^2\log y_i\right]\,,\\
L_2(y_i)&=&\frac{1}{6(1-y_i)^4}\left[-10+43y_i-78y_i^2+49y_i^3-4y_i^4-18y_i^3\log y_i\right]\,.
\eea

\end{appendix}

\end{document}